\documentclass[acmsmall]{acmart}

\AtBeginDocument{%
  \providecommand\BibTeX{{%
    \normalfont B\kern-0.5em{\scshape i\kern-0.25em b}\kern-0.8em\TeX}}}

\setcopyright{acmcopyright}
\copyrightyear{2024}
\acmYear{2024}

\acmJournal{TOSEM}

\usepackage{multirow}
\usepackage{makecell}
\usepackage{pgf-pie}
\usepackage{tikz,stackengine}
\usepackage{caption}
\usepackage{subcaption}
\usepackage{mathrsfs}
\usepackage{titlesec}
\usepackage{enumitem}
\usepackage{bm} 

\titleformat{\subsubsection}
  {\normalfont\normalsize\bfseries}{\thesubsubsection}{1em}{}
\titlespacing*{\subsubsection}{0pt}{3.25ex plus 1ex minus .2ex}{1.5ex plus .2ex}

\usepackage{pgfplots}
\usepackage{pgfplotstable}

\usepackage{changepage}  

\usepackage{graphicx}
\usepackage{multirow}

\usepackage{enumitem}
\usepackage{graphicx}

\usepackage{makecell}

\usepackage[most]{tcolorbox}

\newtcbtheorem{Summary}{\bfseries Summary of RQ}{enhanced,
	coltitle=black,
	top=0.1in,
	attach boxed title to top left=
	{xshift=1.5em,yshift=-\tcboxedtitleheight/2},
	boxed title style={size=small,colback=lightgray}
}{summary}

\newtcolorbox[auto counter]{summary}[1][]{title={\bfseries Summary},enhanced,
	coltitle=black,
	top=0.17in,
	attach boxed title to top left=
	{xshift=1.5em,yshift=-\tcboxedtitleheight/2},
	boxed title style={size=small,colback=lightgray},#1}

\usepackage{mdframed}
\definecolor{shadecolor}{gray}{0.85} 

\mdfdefinestyle{GrayQuoteStyle}{%
  backgroundcolor=shadecolor!80, 
  linewidth=0pt,
  skipbelow=\topskip,
  skipabove=\topskip
}

\usepackage{float}
\usepackage{colortbl}
\usepackage{tcolorbox}
\usepackage{color}

\usepackage{titlesec}
\titlespacing{\section}{0pt}{0.5ex plus 0.2ex minus 0.1ex}{0pt}

\usepackage{array} 
\usepackage{booktabs}
\usepackage{pdflscape}

\usepackage{tikz}
\usepackage{amsmath}

\usepackage{listings}
\usepackage{fancyvrb}
\usepackage{courier}
\usepackage{multicol}
\usepackage{xcolor}
\usepackage{changepage}
\usepackage{adjustbox}
\usepackage{rotating}

\newcommand{\starwithborder}{\huge\textcolor{yellow!90!black}{$\star$}}
\newcommand{\unfilledstar}{\huge\textcolor{black!50}{$\star$}}

\usepackage[most]{tcolorbox}

\begin{document}

\title{LLM-Cure: LLM-based Competitor User Review Analysis for Feature Enhancement}

\author{Maram Assi}
\email{assi.maram@uqam.ca}
\affiliation{
  \institution{Université du Québec à Montréal}
  \city{Montréal}
  \state{Québec}
  \country{Canada}
}

\author{Safwat Hassan}
\email{safwat.hassan@utoronto.ca}
\affiliation{
  \institution{University of Toronto}
  \city{Toronto}
  \state{Ontario}
  \country{Canada}
}

\author{Ying Zou}
\email{ying.zou@queensu.ca}
\affiliation{
  \institution{Queen's University}
  \city{Kingston}
  \state{Ontario}
  \country{Canada}
}


\begin{abstract}
\noindent The exponential growth of the mobile app market underscores the importance of constant innovation and rapid response to user demands. As user satisfaction is paramount to the success of a mobile application (app), developers typically rely on user reviews, which represent user feedback that includes ratings and comments to identify areas for improvement. However, the sheer volume of user reviews poses challenges in manual analysis, necessitating automated approaches.
Existing automated approaches either analyze only the target app’s reviews, neglecting the comparison of similar features to competitors or fail to provide suggestions for feature enhancement. To address these gaps, we propose a \textit{Large Language Model \textbf{(LLM)}-based \textbf{C}ompetitive \textbf{U}ser \textbf{R}eview Analysis for Feature \textbf{E}nhancement)} (\textit{LLM-Cure}), an approach powered by LLMs to automatically generate suggestions for mobile app feature improvements. More specifically, \textit{LLM-Cure} identifies and categorizes features within reviews by applying LLMs. When provided with a complaint in a user review, \textit{LLM-Cure} curates highly rated (4 and 5 stars) reviews in competing apps related to the complaint and proposes potential improvements tailored to the target application. We evaluate \textit{LLM-Cure} on 1,056,739 reviews of 70 popular Android apps. Our evaluation demonstrates that \textit{LLM-Cure} significantly outperforms the state-of-the-art approaches in assigning features to reviews by up to 13\% in F1-score, up to 16\% in recall and up to 11\% in precision. Additionally, \textit{LLM-Cure} demonstrates its capability to provide suggestions for resolving user complaints. We verify the suggestions using the release notes that reflect the changes of features in the target mobile app. \textit{LLM-Cure} achieves a promising average of 73\% of the implementation of the provided suggestions, demonstrating its potential for competitive feature enhancement.
\end{abstract}



\maketitle

\section{Introduction}
\label{sec:introduction}
\noindent The mobile application (app) market is experiencing explosive growth, with global downloads reaching a staggering 257 billion in 2023 \cite{Ceci}. This surge in app adoption has fostered a highly competitive environment among apps within the same categories that offer similar functionalities, i.e., competitors. For instance, WhatsApp\footnote{https://play.google.com/store/apps/details?id=com.whatsapp}, leading the messaging app category with 51 million monthly downloads, competes with at least ten other apps, each having millions of downloads \cite{Laura}. A similar competition is evident in the video conferencing category, where Zoom\footnote{https://play.google.com/store/apps/details?id=us.zoom.videomeetings} and Skype\footnote{https://play.google.com/store/apps/details?id=com.skype.raider} are key players\cite{defilippis2022impact, Flaherty_2020}. During the pandemic, Zoom swiftly adapted to user demands by optimizing its platform for large-scale virtual meetings and enhancing security features, while Skype struggled to keep pace \cite{Stokel_Walker_2020, Rigg_Myle_2021}. To stay relevant and competitive, developers must rapidly respond to user needs. User reviews contain rich information, such as feedback, dissatisfaction and suggestions regarding the user experience of the usage of mobile apps. These reviews offer developers critical insights into areas for improvement and opportunities for feature enhancements \cite{6636712, wei2023enhancing, 8606261, 8057860, 9066126}. To stay competitive, developers need to learn from their competitors' behaviours to maintain a competitive edge \cite{ShahSP19}. Competitor user review analysis involves comparing user feedback, ratings, and reviews of competing mobile applications to identify strengths and weaknesses relative to competitors. By analyzing user reviews from competing apps, developers can uncover insights into features that address unmet needs, potentially giving their app a significant advantage.

Given the sheer volume of reviews \cite{8804432}, it is challenging to analyze user reviews manually. Practitioners need an automated feedback analysis process \cite{9066126}. Hence, researchers propose automated approaches to filter informative reviews \cite{9952173, DBLP:conf/icse/ChenLHXZ14}, summarize user reviews \cite{fu2013people, DBLP:conf/kbse/VuNPN15} and extract features from user reviews \cite{DBLP:conf/refsq/DalpiazP19, DBLP:conf/kbse/VuNPN15, DBLP:conf/msr/IacobH13, 8048887, DBLP:conf/refsq/ShahSP19}. While existing research has been conducted on automated user review analysis, only a limited number of studies focus on competitor user review analysis, where reviews across competing apps are compared
\cite{assi2021featcompare, Quim_2024, Li, ShahSP19, DBLP:conf/sigsoft/ShahSP16, WANG2022118095, 9825826}. Recent advancements in Large Language Models (LLMs) have demonstrated their capabilities across various natural language processing (NLP) tasks, including text generation, translation, summarization, and question answering \cite{minaee2024large}. Although LLMs offer promising applications for mobile app review analysis, current LLM-based research primarily focuses on tasks, such as sentiment analysis \cite{ROUMELIOTIS2024100056, zhang2023sentiment}, aspect extraction \cite{xu2023limits}, analyzing multilingual reviews \cite{10356483} and accessibility-related reviews \cite{10.1145/3638067.3638081}.

Researchers have explored various approaches to conduct competitor user review analysis \cite{DBLP:conf/refsq/DalpiazP19, Li, DBLP:conf/sigsoft/ShahSP16, ShahSP19, assi2021featcompare, LIU2021129}. However, existing work presents some limitations. First, the existing approaches often generate an overwhelming number of fine-grained features \cite{DBLP:conf/refsq/DalpiazP19, DBLP:conf/sigsoft/ShahSP16, ShahSP19} due to comparing the apps' features based on word pairs, making it hard to conduct competitor user review analysis with thousands of features \cite{DBLP:journals/corr/abs-1810-05187}. Second, competitor user review analysis is only conducted by identifying explicit expressions of comparison (e.g., \textit{"Zoom's screen sharing is way smoother than Skype's"}) \cite{Li} and fails to take into consideration implicit insights derived from user reviews. Third, existing work on competitor user review analysis \cite{assi2021featcompare} offers only feature rating comparisons lacking the ability to suggest concrete improvements for specific features based on competitors' user feedback.

To address the limitations of existing work in suggesting feature enhancements using competitor user review analysis, we propose an \textit{\textbf{LLM}-based \textbf{C}ompetitive \textbf{U}ser \textbf{R}eview Analysis for Feature \textbf{E}nhancement (LLM-Cure)}. \textit{LLM-Cure} automatically generates suggestions for mobile app feature improvements by leveraging user feedback on similar features from competitors. \textit{LLM-Cure} operates through two phases. In the first phase, it leverages its large language model capabilities to extract and assign features to user reviews. In the second phase, it curates underperforming features among those identified in the first phase for the target app and suggests potential improvements for specific complaints by leveraging highly rated similar features in competing apps. 

To evaluate the effectiveness of our proposed approach, we conduct an empirical study on 1,056,739 reviews of 70 popular mobile apps from the Google Play store belonging to 7 categories. We evaluate the ability of \textit{LLM-Cure} to (1) accurately assign features to user reviews and (2) offer the developers targeted suggestions for improving the features of their apps based on a specific complaint. We structure our study along by answering the following research questions (RQs):\\

\noindent\textbf{RQ1: How effective can LLMs be in extracting features from user reviews?}
\begin{adjustwidth}{0.8cm}{}
        Automatically extracting features from user reviews provides developers with an efficient way to gain valuable insights into the strengths and weaknesses of mobile app features. In this RQ, we evaluate the ability of \textit{LLM-Cure} to automatically extract features from user reviews. We show that \textit{LLM-Cure} achieves an average F1-score of 85\%, an average recall of 84\% and an average precision of 86\% in assigning features to user reviews, outperforming the state-of-the-art approach by 7\%, 9\% and 4\% in F1-score, recall and precision respectively on average.  \\
 \end{adjustwidth}

\noindent\textbf{RQ2: Can LLMs leverage categorized user reviews to generate suggestions for feature}
\begin{adjustwidth}{0.8cm}{}
     \textbf{improvements?}\\ 
      Automatically generating feature improvement suggestions from competitor user reviews allows developers to address specific complaints and stay competitive in the market. By leveraging the extracted features from user reviews, \textit{LLM-Cure} can pinpoint underperforming features and provide actionable suggestions for enhancement. To validate the suggestions generated by \textit{LLM-Cure}, we cross-reference the suggestions with the release notes, verifying if similar improvements have been implemented in the subsequent releases. We find that 73\% of the suggested enhancements by \textit{LLM-Cure} are implemented in the release notes.\\
 \end{adjustwidth}

The main contributions of our work are as follows:
\begin{enumerate}[label=(\arabic*), align=left, leftmargin=*, itemindent=0pt, itemsep=0pt, topsep=0pt, partopsep=0pt]
    \item We propose an LLM-based approach, \textit{LLM-Cure}, for automatically suggesting feature enhancements through competitor user review analysis.
    \item We develop an efficient method for extracting and assigning features to user reviews, surpassing state-of-the-art approaches by an average of 7\% in F1-score.
    \item We conduct an empirical study on 1,056,739 user reviews of popular apps from the Google Play Store to evaluate the effectiveness of \textit{LLM-Cure} in providing suggestions for feature enhancements. \textit{LLM-Cure} achieves a promising 73\% Suggestions Implementation Rate.\\
\end{enumerate}

\textbf{Paper organization.} The rest of the paper is structured as follows. Section~\ref{sec:background} provides background about LLMs. Section~\ref{sec:approach} presents the overall proposed approach. Section~\ref{sec:experiment} shows our experiments and describes the experimental setups, and results of our research questions. Section~\ref{sec:threats} describes the possible threats of this study. Section ~\ref{sec:related_work} discusses the related work. Lastly, Section~\ref{sec:conclusion} wraps up the study and explores future research directions.

\section{Background}
\label{sec:background}
\noindent\textbf{Large Language Models}. Pre-trained Large Language Models (LLMs) are deep neural networks that have undergone extensive training on large text data that have enabled them to learn complex patterns and structures of language \cite{zhao2023survey}. In the realm of software engineering, LLMs have gained significant attention and adoption for various apps \cite{hou2024large}, including code generation \cite{dong2023selfcollaboration}, code repair \cite{Fan2022ImprovingAG}, and documentation generation \cite{xue2024automated}. Hence, LLMs offer promising avenues for automating software development processes and enhancing developer productivity. Although LLMs are primarily designed for generating text, their output can be influenced by specific instructions communicated through prompts \cite{9908590}. \\

\noindent\textbf{Prompting.} Prompting is a technique used to communicate expectations and guide the LLM's vast knowledge and capabilities towards achieving a specific goal \cite{zhou2023large}. Prompting involves providing instructions to LLMs to guide their generation process and elicit specific types of responses, enforce constraints, or guide the model toward certain stylistic elements. In the software engineering realm, prompting can be used to elicit code snippets, documentation, or other relevant text based on user requirements. \\

\noindent\textbf{In-Context Learning and Few-Shot Learning.} Leveraging pre-trained models, i.e., LLMs, for downstream tasks often requires further fine-tuning on domain-specific labeled data \cite{hu2021lora}. However, fine-tuning LLMs for specific tasks can be computationally expensive and resource-intensive, requiring substantial amounts of task-specific annotated data \cite{brown2020language}. Hence, in-context learning and few-shot learning offer a powerful alternative \cite{dong2023survey}. In-context learning allows the LLM to adapt within a single interaction, using some initial information, i.e., context. Few-shot learning consists of exposing the model to a few examples (i.e., "\textit{few shots}") to make the model effective for a specific task. For instance, for sentiment analysis in user reviews where the goal is to classify reviews as positive or negative, few-shot learning consists of giving the LLM a few examples for each sentiment, i.e., reviews with associated sentiment. The LLM then uses its existing knowledge and these few examples to categorize new reviews based on the context provided.\\

\noindent\textbf{Retrieval Augmented Generation.} LLMs can suffer from hallucination and generate irrelevant or inaccurate responses \cite{huang2023survey}. This can occur due to limitations in their training data or the lack of clear context in the prompt. Retrieval-Augmented Generation (RAG) addresses the issue of hallucination by integrating information retrieval techniques into the generation process \cite{mialon2023augmented}. Thus, the RAG retrieves relevant knowledge from external sources based on the input context, augmenting the model's understanding. For instance, in the realm of user reviews, RAG can be employed to enhance the quality of review summarization. The RAG model augments the user reviews with additional information extracted from external sources, such as sentiment analysis scores, key phrases or keywords, to generate concise and informative summaries of user reviews.

\section{LLM-Cure}
\label{sec:approach}
\begin{figure*}
  \centering
  \includegraphics[width=1.05\textwidth]{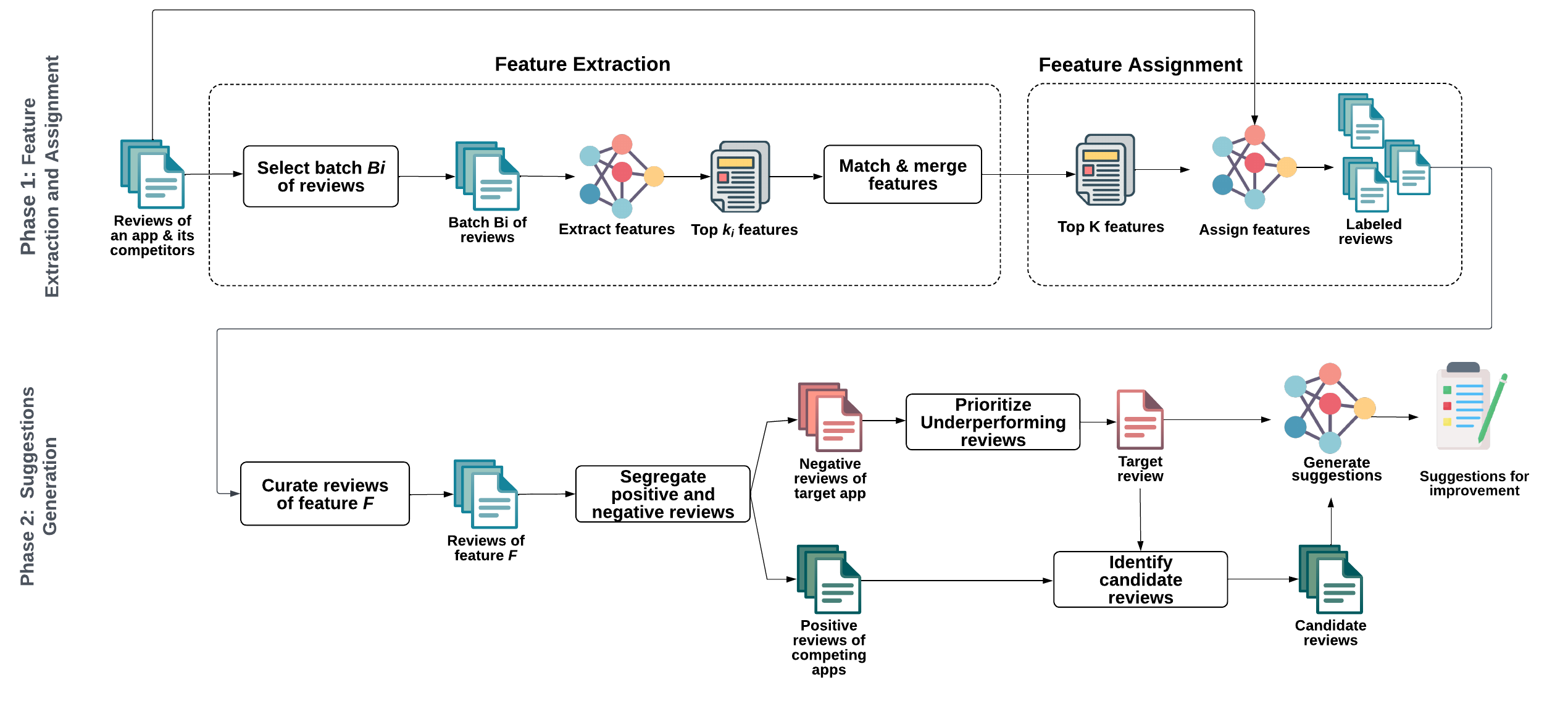} 
  \caption{Overall approach of \textit{LLM-Cure}}
  \label{fig:approach}
\end{figure*}

\noindent\textit{LLM-Cure} is designed to identify user complaints from user reviews and provide suggestions for developers to enhance features that require the developer's attention. More specifically, \textit{LLM-Cure} operates in two distinct phases: (1) \textit{Scalable Feature Extraction and Assignment} that focuses on identifying and assigning features from user reviews of competing apps and (2) \textit{Suggestion Generation with Competitor Reviews} that leverages the extracted features to identify user complaints from a target app and generate suggestions for feature enhancement leveraging the competitor reviews. By incorporating user feedback from competitor reviews, \textit{LLM-Cure} helps developers address user complaints with a competitive edge. Figure \ref{fig:approach} provides an overview of the approach.

\subsection{Scalable Feature Extraction and Assignment}
\label{sub:phase1}
\textbf{Step 1: Extracting Features with Batching and Matching}. This step focuses on identifying the top features that can be summarized by LLMs from an extensive collection of user reviews. However, LLMs have limitations on the amount of context they can process at once. To address this challenge and make our approach scalable, we introduce the so-called \textit{batch-and-match} approach that incrementally extracts the top $k$ features from a large corpus of user reviews. Our approach consists of three processes:\\

\newcommand{\circled}[1]{%
  \tikz[baseline=(char.base)]\node[draw, shape=circle, inner sep=1pt] (char) {#1};%
}

\begin{figure*}
  \centering
  \includegraphics[width=\textwidth]{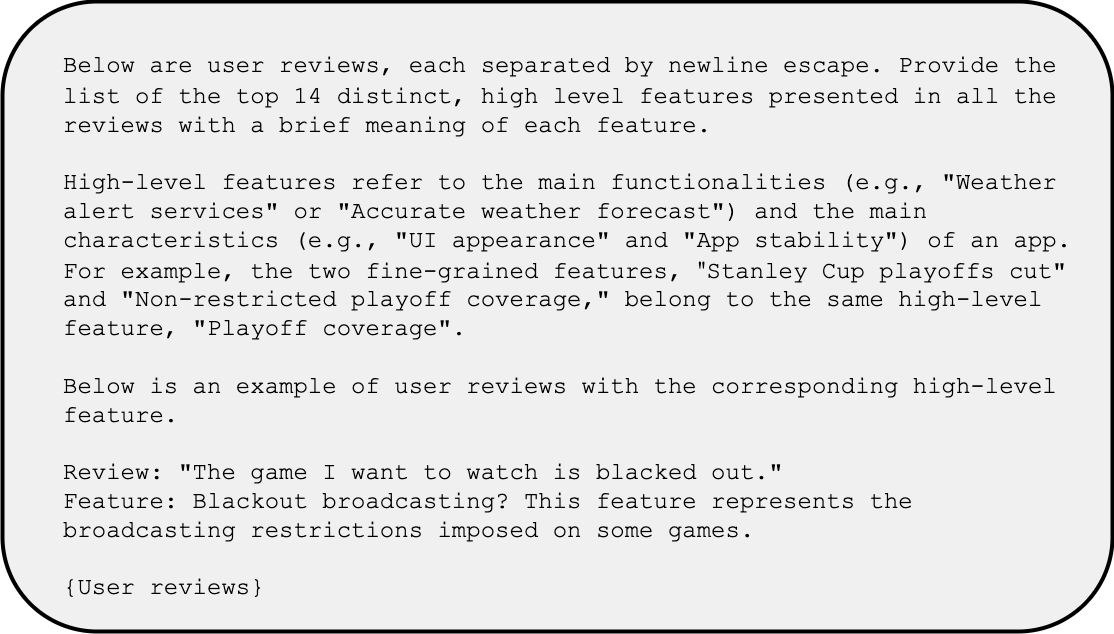} 
  \caption{Template of \textit{LLM-Cure}'s prompt for feature extraction}
  \label{fig:prompt_extract}
\end{figure*}

\noindent \circled{1} \textit{Batching reviews and extracting features.} Batching reviews involves dividing a large volume of reviews into manageable batches for incremental processing and feature extraction using LLMs. First, we shuffle the entire collection of user reviews of a group of competing apps to ensure randomness. Let $R = \{r_1, r_2, ..., r_m\}$ represent the shuffled user reviews where $m$ is the total number of reviews. To efficiently process the large volume of reviews, we split the shuffled reviews $R$ into batches of a predefined size $s$ (e.g., 1,000 reviews) that fit within the LLM context size. We denote the batches as $B = \{B_1, B_2, ..., B_n\}$, where $n$ is the total number of batches. Each batch $B_i$ corresponds to a set of individual reviews $R_i = \{r_i, r_{i+1}, ..., r_{i+s}\}$, with each $r_i$ being a user review in batch $B_i$. We process each batch $B_i$ sequentially using the LLM. For each batch $B_i$, we prompt the LLM to extract the top $k$ features, denoted as $F_i$, from the set of reviews $R_i$. The value of $k$ is a hyper-parameter that can be tuned to optimize the performance of \textit{LLM-Cure} depending on the selected dataset. The \textit{Feature extraction} prompt, illustrated in Figure \ref{fig:prompt_extract}, identifies the top $k$ features for a specific app category. It is structured to encapsulate the task description, define features, include a one-shot example, and present the list of user reviews.\\

\noindent \circled{2} \textit{Matching and merging similar features.} 
Since we process thousands of reviews in batches due to limitations on the LLM's context handling, the extracted features might sometimes use different wording. For instance, in one batch, the LLM might identify \textit{"Advertisements"} as a feature, while another batch might highlight \textit{"In-app Advertisements"}. Both features represent the same functionality extracted from the user reviews. To ensure a non-redundant feature identification, we address this challenge by incrementally combining similar features after processing each batch $B_i$. $F_i$ represents the features extracted from batch $B_i$. We define $M_i$ as the set of matched and merged features until batch $B_i$. The merging process starts with the top $K_{1}$ features extracted from the first batch $B_1$. For the first batch, $M_1$ = $F_1$. For subsequent batches ($B_i$, $i$ > 1), we compare features in $F_i$ with $M_{i-1}$ features already identified from the previous batch $B_{i-1}$ and match then merge the similar features. Word embeddings and cosine similarity are employed to achieve the merging. Word embeddings represent the features in a high-dimensional vector space, capturing their semantic meaning \cite{Bengio2003ANP}. Cosine similarity, a metric for measuring similarity between vectors, is then calculated between the embedding vectors of features $F_i$ and $M_{i-1}$ from different batches. Features exceeding a predefined similarity threshold $tr$ in cosine similarity are considered highly similar and subsequently matched and merged. The similarity threshold $tr$  is a hyper-parameter. Therefore, we experiment with thresholds ranging from 0.7 to 0.85 on a validation set and choose the value that leads to the highest precision. This incremental process continues with each new batch $B_i$, merging similar features from $F_i$ with the existing merged set $M_{i-1}$ leading to a unique set of features $M_{i}$.\\

\noindent \circled{3} \textit{Verifying convergence and stabilizing features.} The challenge in this step is to determine when the incremental \textit{batch-and-match} process has sufficiently captured the top $k$ features, avoiding unnecessary iterations that would consume additional processing time and computational resources to process the entire volume of user reviews. We address this by defining a \textit{convergence threshold} based on the stability of the top $k$ features over a specified number of consecutive iterations $N$. For instance, the system starts with the initial merged set ($M_1$). The batch processing continues until the merged set $M_{j}$ where the merged sets remain unchanged across the last $N$ iterations (from $M_j$ up to $M_{j-N}$). For example, assuming the \textit{convergence threshold} is set to 3. \textit{LLM-Cure} checks if the top $k$ features identified have remained stable for the last 3 batches. This stability indicates that we have captured the dominant features in the reviews $R$, and further processing would probably not yield new features. The \textit{convergence threshold} $N$ is a hyper-parameter. Therefore, we experiment with thresholds equal to 3, 5, and 7 on a validation set. Following this convergence step, we obtain the final set of top $k$ features extracted from the review batches.\\

\begin{figure*}
  \centering
  \includegraphics[width=\textwidth]{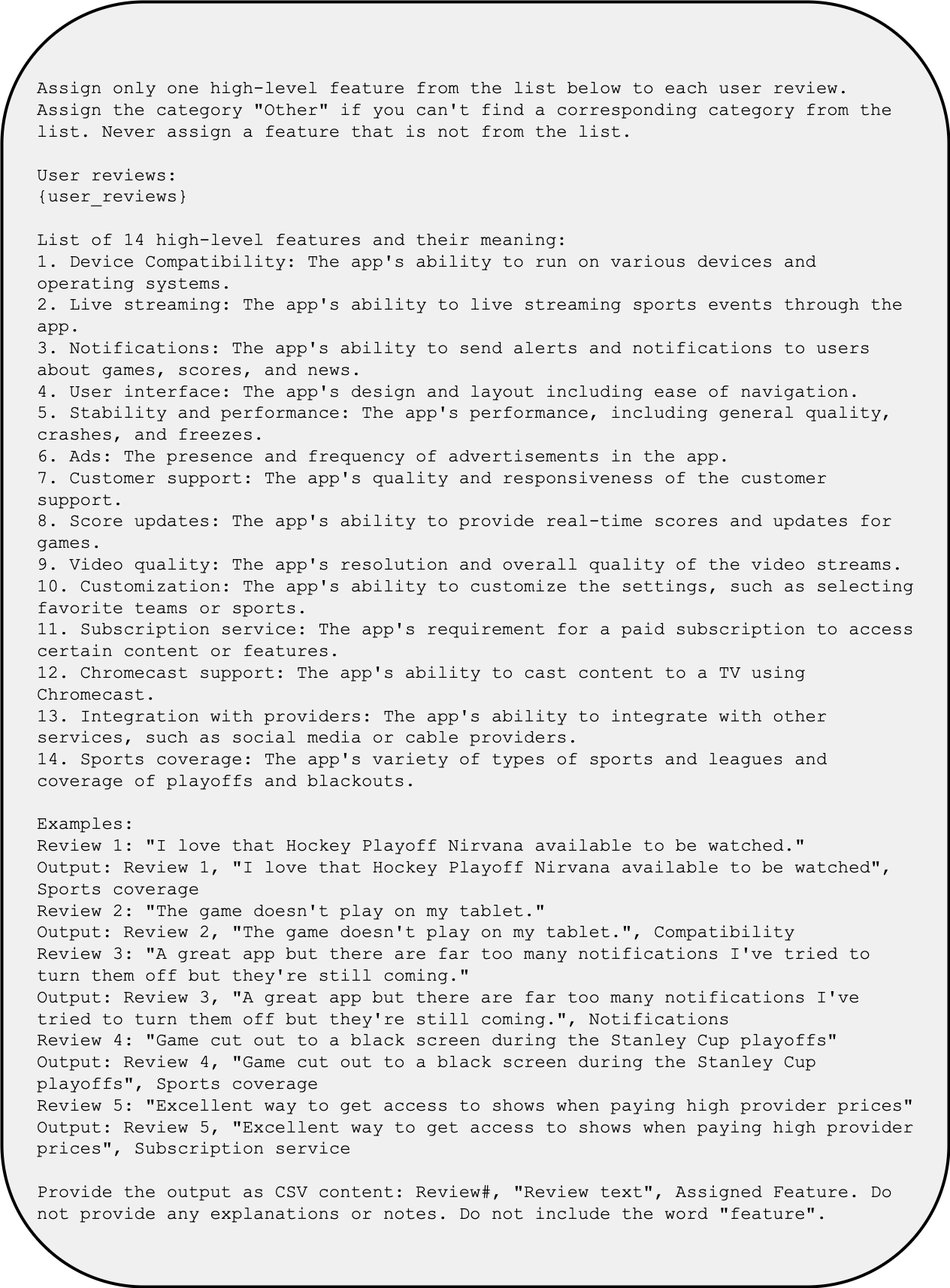} 
  \caption{Example of \textit{LLM-Cure}'s prompt for feature assignment for the Sports News Category}
  \label{fig:prompt_assign}
\end{figure*}

\noindent \textbf{Step 2: Assigning Features to Reviews.} The prior research on the dataset \cite{assi2021featcompare} used in our approach demonstrates that only 8.6\% of the user reviews contain multiple features and that multi-labeling does not lead to a significant impact on the results. We leverage prior findings to task a language model to assign one feature to the reviews by constructing a \textit{Feature assignment} prompt. The \textit{Feature assignment} prompt incorporates the task description, the extracted $k$ features with their brief meaning, five few-shot examples demonstrating feature assignment to user reviews, and the list of user reviews to be classified. Figure \ref{fig:prompt_assign} shows the \textit{Feature assignment} prompt for the Sports News category. As a result, each user review $r_i$ is associated with one designated feature, building the groundwork for targeted analysis and feature enhancement suggestions generation in the second phase.

\subsection{Suggestion Generation with Competitor Reviews}
\label{subsec:phase2}

\noindent Prior research \cite{10.1145/2487575.2488202, 9794110} shows that negative reviews, those with 1 or 2-star ratings, are particularly interesting to developers as they often contain valuable insights regarding feature complaints and areas for enhancement. Conversely, positive reviews, typically rated 4 or 5 stars, offer detailed descriptions of features and positive user experience \cite{7320414}. These positive reviews are valuable resources as they often showcase successful implementations of similar features and offer potential solutions to address user complaints. Additionally, prior research \cite{AralikatteSGM18, sanger-etal-2016-scare} indicates that 3-star ratings are typically viewed as neutral, or often encompass both praise and criticism of the app features.  

In Phase 2, we leverage the user feedback summarized from competitors’ positive reviews to provide suggestions for the target apps to improve the features associated with negative reviews. Our proposed method uses an RAG approach to dynamically construct prompts for the LLM that are augmented with relevant positive user reviews from competing apps. By analyzing these positive reviews, the LLM can identify successful implementations and suggest potential solutions to user complaints within the target application. \textit{LLM-Cure} generates suggestions based on the following five distinct steps.\\

\noindent\textbf{Step 1: Curating Popular Underperforming Features.} An underperforming feature is defined as one that has the largest percentage of negative reviews. Identifying underperforming features is crucial for developers to prioritize areas for improvement and focus on features with the highest percentage of negative reviews to address user dissatisfaction better. We calculate the \textit{Underperforming Feature Score (UFS)} for each feature by determining the percentage of negative reviews associated with it. The formula for \textit{UFS} for a particular feature $F$ is:\\

\begin{equation}
\text{UFS}_F = \frac{\text{Number of Negative Reviews}_F}{\sum_{i=1}^{k} \text{Number of Negative Reviews}_i} \times 100
\end{equation}\\

\noindent \parbox{\linewidth}{ Where the \(\text{Number of Negative Reviews}_F\) denotes the number of negative reviews associated with feature $F$, and \(\sum_{i=1}^{k} \text{Number of Negative Reviews}_i\) denotes the total number of negative reviews across all $k$ features. Sorting features in descending order of their UFS prioritizes those with the highest percentage of negative reviews. This allows developers to focus on features most frequently associated with user dissatisfaction.\\}

\noindent \textbf{Step 2: Segregating Positive and Negative Reviews.} In this step, for a selected underperforming feature $F$, we select the negative reviews rated 1 and 2 stars of the target app. Concurrently, we also curate positive reviews rated 4 or 5 stars of the same feature $F$ from competitor apps. We exclude reviews associated with a 3-star rating.\\

\noindent \textbf{Step 3: Prioritizing Complaint-Rich Negative Reviews.} Not all reviews contain the same amount of details. Some might express generic dissatisfaction, while others delve deeper and provide specific details about the issues encountered with the feature. In this step, we aim to guide developers towards the most informative negative reviews, i.e., complaint-rich, for a specific underperforming feature $F$. To accomplish this, we implement a ranking mechanism on the negative reviews associated with the target app, utilizing the TF-IDF (term frequency-inverse document frequency) score \cite{Ramos}. TF-IDF analyzes the importance of words within a document (in this case, a user review) relative to their occurrence across the entire dataset of reviews. For a specific feature $F$ of a target app, we use TF-IDF to calculate a score for each negative review. We sum up the TF-IDF scores of all words in a review to get a final TF-IDF score for the entire review. Higher TF-IDF scores indicate that the user review contains terms that are both frequent in the review and relatively unique across the entire dataset, suggesting detailed and specific feedback. This score reflects the review's richness in terms of feature-specific complaints. We select the top n negative reviews based on the calculated TF-IDF scores. These selected reviews $R=\{r_1, r_2, ..., r_k\}$ guide developers, directing their attention towards negative feedback rich in informative content regarding feature complaints specific to the target app.\\

\noindent \textbf{Step 4: Identifying Candidate Reviews for Relevant Solutions.} This step aims to identify potential recommendations from competitors' user reviews that might address the complaint in a specific negative review. This step consists of four processes:\\

\noindent \circled{1} \textit{Selecting a negative review.} We start by picking one of the top n negative reviews (e.g., $r_1$) identified in Step 3.

\noindent \circled{2} \textit{Selecting candidate reviews from competitors.} We look at positive (4 and 5-star) reviews $P_F$ for the same feature $F$ selected in Step 2. To mimic a practical environment where developers might only have access to historical data, we further filter out the candidate positive reviews to include only those with a post date equal to or before the selected target complaint review $r_1$. These reviews represent the candidates for finding suggestions for improvement. 

\noindent \circled{3} \textit{Creating vector embeddings.} We employ an LLM-based word embedding technique to convert the reviews into vector representations to compare them based on their semantic meaning. Using the same embedding model, we generate the vector representations for the selected negative review $r_1$ and all positive reviews $P_F$. Let \( V_{r_1} \) represent the vector representation of the selected negative review \( r_1 \), and \( V_{p_i} \) denote the vector representation of a positive review \( p_i \).

\noindent \circled{4} \textit{Finding similar reviews.} Since all reviews are represented in the same vector space, we leverage cosine similarity to compare the vectors. The cosine similarity \( \text{sim}(V_{r_1}, V_{p_i}) \) between the vector representations of the negative review and each positive review is calculated as:\\
\begin{equation}
\text{sim}(V_{r_1}, V_{p_i}) = \frac{V_{r_1} \cdot V_{p_i}}{\|V_{r_1}\| \cdot \|V_{p_i}\|}
\end{equation}\\
\noindent We rank the positive reviews based on their similarity, and we identify a sample \( P \) of positive reviews that exhibit the highest similarity to the negative review. These selected positive reviews represent instances where users discuss similar feature characteristics but in a positive context.\\

\begin{figure*}
  \centering
  \includegraphics[width=\textwidth]{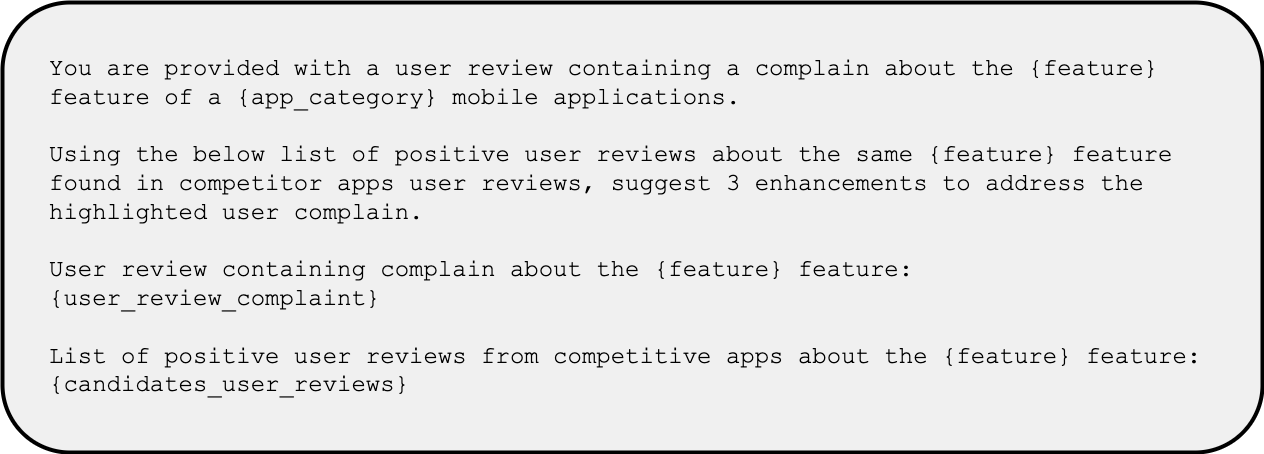} 
  \caption{Template of \textit{LLM-Cure}'s prompt for featue improvement suggestions}
  \label{fig:prompt_suggest}
\end{figure*}

\noindent \textbf{Step 5: Prompting Suggestions Using RAG}. This step revolves around constructing the prompt for instructing the LLM to generate the relevant suggestions. Specifically, we design RAG-based prompts by leveraging the positive reviews identified in Step 4 to provide contextual guidance to the LLM regarding where to draw suggestions. Then, we instruct the LLM to suggest top N unique and constructive recommendations to enhance the feature $F$, discussed in review $r_1$, using the provided sample of positive reviews identified in Step 4. Figure \ref{fig:prompt_suggest} illustrates the template for the \textit{Improvement Suggestions} prompt. This RAG-based prompt provides developers with suggestions derived from positive user experiences, facilitating targeted improvements to address user concerns effectively.

\subsection{Implementation of LLM-Cure}
\label{sub:implementation}
\textbf{LLM Choice.} We select \textit{Mixtral-8x7B-Instruct-v0.1}\footnote{https://mistral.ai/news/mixtral-of-experts/} LLM to conduct our experiments. \textit{Mixtral-8x7B-Instruct-v0.1} is a high-performing, open-weight Sparse Mixture of Experts model. We choose this model as it balances cost with performance. It has been demonstrated that it surpasses open source models, including \textit{Llama 2 70B}\footnote{https://llama.meta.com/llama2/} while achieving 6x faster inference, and it matches \textit{GPT-3.5}\footnote{https://openai.com/chatgpt/} performance on standard tasks \cite{AI_2024}. Being open-source and free, Mixtral allows other researchers to easily access, understand, and adapt our work. We employ Python scripts to facilitate loading the \textit{Mixtral-8x7B-Instruct-v0.1} model from the Hugging Face Hub\footnote{   https://huggingface.co/mistralai/Mixtral-8x7B-Instruct-v0.1}.\\

\noindent \textbf{Embedding Model Choice.}
To ensure consistency within our approach, \textit{LLM-Cure} employs the \textit{mistral-embed}\footnote{https://docs.mistral.ai/api/} word embedding model from Mistral AI during the process that requires word embedding. Specifically, we used it in the \textit{Matching Similar Features} process of phase 1 and in the \textit{Identifying Candidate Reviews for Relevant Solutions} step of phase 2. We leveraged the Mistral API to retrieve text embeddings efficiently.\\

\noindent \textbf{Text Preprocessing.} Prior to feeding text inputs into the \textit{mistral-embed} model, we conducted standard text normalization processes adopted in previous work \cite{10.1145/3593802, 10499727} to enhance the quality of the input data by applying tokenization, removal of stop words, stemming, and spell-checking. We employed the SpellChecker\footnote{https://pypi.org/project/pyspellchecker/} along with the nltk\footnote{https://pypi.org/project/nltk/} libraries in Python for these preprocessing steps.

\section{Experimental Design}
\label{sec:experiment}
\noindent We aim to evaluate the ability of \textit{LLM-Cure} to (1) accurately assign features to user reviews and (2) provide the developers with suggestions to improve the features of their apps given a specific complaint. In this section, we describe the experimental setup and discuss the results of our investigation.

\begin{table}[h]
\centering
\caption{Descriptive statistics of 7 competing apps categories}
\label{tab:keywords}
\begin{tabular}{lcc}
\toprule
\textbf{App category} & \textbf{Number of apps} & \textbf{Number of reviews} \\ \midrule
Free call & 10 & 291,034\\
Weather & 10 & 276,941\\
SMS & 10 & 264,926 \\
Bible & 10 & 64,417\\
Music player & 10 & 60,685\\
Sports news & 10 & 57,582\\
Cooking recipe & 10 & 41,154\\

\midrule
\textbf{Total} &  70 & 1,056,739 \\
\bottomrule
\end{tabular}
\end{table}

\subsection{Dataset}
\label{dataset}
We employ the same dataset utilized by a previous work \cite{assi2021featcompare}. The original dataset comprises 20 categories of competing apps selected from the top 2,000 popular free-to-download apps from the Google Play Store. The selected apps span diverse categories, e.g., \textit{Navigation}, \textit{Weather}, \textit{Browser}, \textit{FreeCall} and \textit{Dating}, and each category includes a sufficient number of competing apps (e.g., 8 to 10 competing apps) to facilitate competitor user review analysis. To evaluate our approach against the baselines, we use the same five categories used by the baselines \cite{assi2021featcompare} to evaluate the precision, namely \textit{Weather, SMS, Bible, Music Player}, and \textit{Sports news}. Similar to previous work, we use the \textit{Free call} and \textit{Cooking recipe} categories for hyper-parameter tuning. Specifically, for each category, we utilize the same statistically representative sample of reviews, i.e., 96 user reviews per category, resulting in a total of 672 ground truth user reviews to evaluate our approach. Table \ref{tab:keywords} summarizes the descriptive statistics of the 70 selected competing app categories. The user reviews of each app are also available, and each user review records the title of the user review, detailed comment, user rating and the posting date. Additionally, we have access to the release notes of the apps which allows us to gain information about the features and updates introduced in each app version. 
 
\subsection{Research Questions}
\subsubsection{RQ1: How effective can LLMs be in extracting features from user reviews?}
In our work, \textit{LLM-Cure} suggests enhancements to developers to improve their app features by identifying features from user reviews. Therefore, we want to evaluate the capabilities of the LLM in automatically extracting meaningful features from user reviews to understand the feasibility and potential of \textit{LLM-Cure} for real-world applications.\\

\noindent \textbf{Evaluation Metrics.} 
We assess the performance of \textit{LLM-Cure} by comparing predicted features by the LLM against the ground truth. We employ three key metrics: 1) True Positives (TP) representing the correctly predicted features, 2) False Positives (FP) representing the number of falsely predicted features, and 3) False Negatives (FN) representing features present in reviews but not predicted by \textit{LLM-Cure}. We adopt the precision, recall and F1-score as evaluation metrics, and we calculate them as follows:

\begin{gather}
\label{equ:precision}
    Precision = \frac{TP}{TP + FP} \\[6pt] 
\label{equ:recall}
    Recall = \frac{TP}{TP + FN} \\[6pt] 
\label{equ:f_Score}
    F_1\textit{-}Score = 2* \frac{Precision * Recall}{Precision + Recall}
\end{gather}\\

To evaluate the performance metrics of feature extraction, the first and second authors, as two independent annotators, manually label 672 testing reviews. The Cohen’s Kappa agreement score \cite{Cohen} is computed on the annotated testing reviews, yielding a high score of 0.82, indicative of a high level of agreement.\\

\noindent\textbf{Experimental Setup.} \textit{LLM-Cure} has three hyper-parameters: $k$ the number of the features, the \textit{similarity threshold} and the \textit{convergence threshold}. Previous work that uses the dataset demonstrates that 14 leads to the best results when set as the number of features. Therefore, we set $k$ as 14, aligning with previous work on this dataset \cite{assi2021featcompare}. To set the \textit{similarity threshold} and the \textit{convergence threshold}, we conduct experiments on the two validation sets, i.e., \textit{"Recipe cooking"} and \textit{"Free Call"} app categories. The results
indicate that a \textit{similarity threshold} of 0.75 coupled with a \textit{convergence threshold} of 5 yields the highest precision. Therefore, we adopt these hyper-parameter values for the testing set. In addition, to prevent exceeding the limited context size of the LLM, we select a batch size of 1,000 reviews. This batch size is determined based on the varying lengths of individual reviews. Through experimentation, we have found that 1,000 reviews is optimal, as it ensures that the total token count of each batch does not exceed the context window of the selected Mistral model. \\

\noindent\textbf{Baselines.} To assess the efficacy of \textit{LLM-Cure} in automatically identifying and assigning features to user reviews, we compare its performance against existing baselines FeatCompare\cite{assi2021featcompare} and Attention-based Aspect Extraction (ABAE) \cite{he-etal-2017-unsupervised}, using the ground truth of 480 labeled reviews. 
In addition, we include a baseline called \textit{LLM-Basic}. Similar to \textit{LLM-Cure}, in \textit{LLM-Basic}, we prompt the LLM to extract features from user reviews. However, in this baseline, we do not include the incremental process of batching, i.e., \textit{batch-and-match}, used in \textit{LLM-Cure}. Instead, we select a statistical sample of reviews from the set of all reviews that fits the context size of the LLM (i.e., 1,000 reviews) and use the same prompt used for \textit{LLM-Cure}. Our goal is to verify that the incremental process adds value to the extraction and improves the performance.\\

\noindent \textbf{Prompt Construction.} To construct the prompts, we adhere to the template provided by Mistral for our selected model\footnote{https://huggingface.co/mistralai/Mixtral-8x7B-Instruct-v0.1}. \textit{LLM-Cure} leverages two distinct prompts for Phase 1: the \textit{Feature extraction} prompt and \textit{Feature assignment} prompt. The \textit{Feature extraction} prompt is illustrated in Figure \ref{fig:prompt_extract} and the \textit{Feature assignment} in Figure \ref{fig:prompt_assign}. The prompts are also available in our research artifact.\\

\noindent\textbf{Results.} \textbf{\textit{LLM-Cure} is capable of identifying and assigning features with high F1-score, recall and precision}. Table \ref{tab:extracted_features_part1} and Table \ref{tab:extracted_features_part2} show the fourteen features extracted for the five testing apps categories. Across the five testing app groups, \textit{LLM-Cure} exhibits F1-scores ranging from 80\% to 91\%, with precision between 81\% and 92\% and recall between 80\% and 90\%. These findings

\begin{table}[H]
\centering
\caption{Top fourteen features extracted from the user reviews of three sample app categories (Part 1)}
\label{tab:extracted_features_part1}
\resizebox{\columnwidth}{!}{
\begin{tabular}{ll}
\toprule
\textbf{Feature} & \textbf{Description}\\
\midrule
\multicolumn{2}{c}{\textbf{Weather}} \\
\midrule
Accuracy & App's ability to provide accurate weather forecasts\\
Radar & App's radar and map and visualization features\\
Weather Forecast & App's hourly and daily forecasts\\ 
Additional Features & App's additional features, such as pollen counts and UV index\\
Notifications & App's ability to send notifications for weather alerts\\
Customization & App's ability to be customized, i.e., adding multiple locations\\
Battery Usage & App's impact on the device battery life\\
Customer Support & Users' experiences with the app's customer support\\
Ease of Use & App's ease of use including widgets and how to navigate\\
Ads & App's use of ads, including how intrusive they are\\
Device Compatibility & App's compatibility with different devices and systems\\
Design and Layout & App's design and layout\\
Performance & App's stability, including how often it crashes or freezes\\
Updates & App's frequency and quality of updates\\
\midrule
\multicolumn{2}{c}{\textbf{SMS}} \\
\midrule
Group Messaging & App's ability to send messages to multiple recipients at once\\
Dual SIM Support & App's ability to support dual SIM devices\\
Account Authentication & App's ability to identify and authenticate a user's account\\
Customization & Ability to customize the app's appearance and/or functionality\\
Spam Identification & App's ability to identify and filter out spam messages\\
MMS Support & App's ability to send and receive multimedia messages\\
Integration with Other Apps & App's ability to integrate with other apps\\
In-app Ads & Presence of advertisements within the app\\
Notifications & App's ability to send notifications for incoming messages\\
Backup and Restore & App's ability to back up and restore messages and settings\\
User Interface (UI) & App's design and layout\\
Call Blocking & App's ability to block unwanted calls\\
Caller Identification & App's ability to identify the caller's name and/or location\\
Privacy & App's ability to protect the user's privacy\\
\midrule
\multicolumn{2}{c}{\textbf{Sport News}} \\
\midrule
Device Compatibility & App's ability to run on various devices and systems\\
Live Streaming & App's ability to live stream sports events\\
Score Updates & App's ability to provide real-time scores and updates for games\\
Sports Coverage & App's variety of types of sports and coverage\\
Notifications & App's ability to send alerts and notifications\\
User Interface & App's design and layout including ease of navigation\\
Chromecast Support & App's ability to cast content to a TV using Chromecast\\
Performance & App's performance, including general quality and crashes\\
Ads & The presence and frequency of advertisements in the app\\
Customer Support & App's quality and responsiveness of the customer support\\
Video Quality & App's resolution and overall quality of the video streams\\
Customization & App's ability to customize the settings\\
Subscription Service & App's requirement for a paid subscription\\
Providers Integration & App's ability to integrate with other services (cable providers)\\
\end{tabular}
}
\end{table}

\begin{table}[H]
\centering
\caption{Top fourteen features extracted from the user reviews of three sample app categories (Part 2)}
\label{tab:extracted_features_part2}
\resizebox{\columnwidth}{!}{
\begin{tabular}{ll}
\toprule
\textbf{Feature} & \textbf{Description}\\
\midrule
\multicolumn{2}{c}{\textbf{Music Player}} \\
\midrule
Playlist Management & App's features for creating, editing, and organizing playlists\\
Lyrics Integration & App's ability to display lyrics in real-time and for offline use\\
Equalizer \& Sound Adjustment & App's options for sound customization\\
Ad-Free Experience & App's option to remove ads\\
Search & App's ability to allow users to search for a specific song\\
Volume Leveler & App's ability to adjust songs' volume to a common level\\
Shuffle & App's ability to shuffle songs in a playlist or library\\
Sleep Timer & App's option to automatically stop playback\\
Sound Quality & App's overall sound quality, including volume and bass\\
User Interface & App's design and layout (e.g., theme customization)\\
Notifications & App's ability to send notifications based on user's preference\\
Offline Mode & App's ability to download songs for offline playback\\
Chromecast/AirPlay Support & App's wireless streaming with Chromecast or AirPlay\\
Download & App's ability to download music for offline listening\\
\midrule
\multicolumn{2}{c}{\textbf{Bible}} \\
\midrule
Bible Versions & The ability to switch between different translations of the Bible\\
Daily Verses & Daily verse for meditation and reflection\\
Reading Plans & Guided plans for reading the Bible over a set period\\
Social Features & The ability to share verses or reflections with users\\
Highlighting \& Bookmarking & The ability to mark specific verses for future reference\\
Search Function & The ability to search for specific verses or topics in the Bible\\
Audio Feature & The option to listen to the Bible being read aloud\\
Customization & App's ability to personalize with different themes and fonts\\
Offline Access & Offline Access and download portions of the Bible\\
Devotionals & Pre-written devotionals on various topics\\
User Experience & App's layout design and user experience\\
In-App Purchases & App's option to buy extra features or content\\
Notifications & App's ability to send reading reminders\\
Comparison Feature & App's ability for comparison of different bible translations\\
\midrule
\end{tabular}
}
\end{table}

underscore the capability of LLMs to extract features from user reviews without requiring manual annotation. 

\textbf{LLM-Cure significantly outperforms LLM-Basic, FeatCompare and ABAE baselines across the testing apps.} Table~\ref{tab:accuracy_res_test} shows that on average, \textit{LLM-Cure} achieves a 7\% improvement in F1-score, a 9\% improvement in recall and a 4\% in precision as compared to the baselines. To quantitatively assess these differences, we conducted paired t-tests. A paired t-test \cite{Mishra2019} is a statistical method used to determine whether there is a significant difference between the means of two related groups. In our case, these groups are the performance metrics of LLM-Cure and FeatCompare, the best performing baseline. Our findings indicate that LLM-Cure significantly outperforms FeatCompare in both F1-score and precision. Specifically, the paired t-test for F1-score yielded a t-statistic of 3.723 with a p-value of 0.02, confirming a statistically significant difference. Similarly, the paired t-test for precision resulted in a t-statistic of 4.784 and a p-value of 0.009, further underscoring LLM-Cure's superior performance.

\textbf{The \textit{batch-and-match} process of \textit{LLM-Cure} improves the performance of feature extraction.} \textit{LLM-Basic}, which only processes a single batch of reviews, achieves lower performance compared to \textit{LLM-Cure}. These results highlight the benefit of \textit{LLM-Cure's} incremental processing, \textit{batch-and-match}, and its ability to extract features more effectively and with higher performance (e.g., F1-score). Instead of processing the entire set of user reviews, \textit{LLM-Cure} processes only a fraction of the total reviews. Table \ref{tab:rq1_batch} illustrates the percentage of user reviews needed by \textit{LLM-Cure} to achieve convergence and extract the features. \textit{LLM-Cure} outperforms all baseline methods while processing only between 3\% and 30\% of user reviews, achieving feature saturation without the need for processing the entire dataset.

\textbf{\textit{LLM-Cure} performs consistently across different sentiment categories.} To further investigate whether \textit{LLM-Cure} classifies positive versus negative reviews with different precision, we conduct an analysis based on the sentiment categories. We find that positive and negative reviews present similar results across categories, with at most 3\% of differences. The average precision for positive reviews across these categories is 85.69\%, while for negative reviews, it is 85.78\%. These findings indicate that there is no significant difference in classification F1-score between positive and negative reviews, demonstrating that our approach is not sensitive to the sentiment of reviews.

\begin{table}[H]
\centering
\caption{The number of batches and percentage of reviews required to extract the features using LLM-Cure}
\label{tab:rq1_batch}
\begin{tabular}{lcc}
\toprule
\textbf{App category} & \textbf{Number of batches} & \textbf{Percentage of reviews}\\ \midrule
Weather & 7 & 3\%\\
Bible & 4 & 6\%\\
SMS & 24 & 9\% \\
Sport News & 8 & 14\%\\
Music player & 19 & 31\%\\
\bottomrule
\end{tabular}
\end{table}

\begin{table}[H]
\centering
\caption{Performance comparison of \textit{LLM-Cure} and baselines on five testing app groups in features assignment. `P' denotes Precision, and `R' denotes Recall and `F\textsubscript{1}' denotes F1-score}
\label{tab:accuracy_res_test}
\begin{tabular}{l|p{0.15cm}p{.15cm}p{0.15cm}|p{0.15cm}p{.15cm}p{0.15cm}|p{0.3cm}p{.25cm}p{0.2cm}|p{0.15cm}p{.15cm}p{0.15cm}
}
\toprule
\textbf{App} & \multicolumn{3}{c|}{\textbf{LLM-Cure}} & \multicolumn{3}{c|}{\textbf{LLM-Basic}} & \multicolumn{3}{c|}{\textbf{FeatCompare}} & \multicolumn{3}{c}{\textbf{ABAE}}\\

\textbf{Category} & P & R & $F_1$ & P & R & $F_1$ & P & R & $F_1$ & P & R & $F_1$\\

\midrule

Weather & \textbf{92} & \textbf{90} & \textbf{91} & 80 & 77 & 79 & 81 & 74 & 78 & 64 & 64 & 67 \\
Sports News & 81 & \textbf{80} & \textbf{80} &  67 & 65 & 66 & \textbf{82} & 75 & 78 & 71 & 65 & 68 \\
Bible & \textbf{83} & \textbf{83} & \textbf{83}  & 78 & 75 & 77  & 81 & 77 & 79 & 72 & 69 & 70 \\
SMS & \textbf{86} & \textbf{83} & \textbf{85}  & 77 & 76 & 77 & 80 & 74 & 77 &67 & 62 & 64\\
Music Player & \textbf{86} & \textbf{86} & \textbf{86}  & 82 & 82 & 82 &  79 & 75 & 77 & 70 & 66 & 68 \\
\midrule
Average & \textbf{86} & \textbf{84} & \textbf{85}  & 77 & 75 & 76 &  82 & 75 & 78 & 68 & 65 & 68 \\
\bottomrule
\end{tabular}
\end{table}

\begin{Summary}{}{firstsummary}
\textit{LLM-Cure} achieves promising F1-scores ranging from 80\% to 91\% across the five test sets, demonstrating its effectiveness in analyzing user reviews without manual data labeling. The obtained F1-scores surpass the performance of the baselines by an average of 7\%. The \textit{batch-and-match} process enables LLM-Cure to achieve a high F1-score with a substantial reduction in the required user review data. \end{Summary}

\subsubsection{RQ2: Can LLMs leverage categorized user reviews to generate suggestions for feature improvements?}
\label{sec:rq2}
RQ2 investigates whether \textit{LLM-Cure} can leverage categorized user reviews of competitor apps to generate suggestions for feature improvements. Analyzing competitors' positive user reviews allows developers to identify successful features and user preferences across the market, ensuring their app remains competitive and relevant. By incorporating these insights from competitor reviews, \textit{LLM-Cure} empowers developers to make data-driven decisions about feature enhancement, prioritize user needs, and ultimately create a more competitive app.\\

\noindent \textbf{Suggestions validation.} To assess the relevance of the suggestions provided by \textit{LLM-Cure}, we conduct a retrospective investigation at the app release level. Specifically, we examine the release notes of future releases of the target app following the date of the user review containing the complaint and calculate the \textit{Suggestions Implementation Rate (SIR)}. We define the \textit{SIR} the number of suggestions by \textit{LLM-Cure} matched in the release notes divided by the total number of suggestions provided as follows:
\begin{equation}
\label{eq:sir}
\text{SIR} = \frac{\text{Number of Suggestions Matched in Release Notes}}{\text{Total Number of Suggestions by LLM-Cure}}
\end{equation}\\

\noindent \textbf{Experimental Setup.} We randomly select three categories to evaluate the feature improvement suggestions. From each category, We select the apps with a substantial number of informative release notes to ensure that we have a rich data source for conducting the manual suggestion validation. Specifically, we choose \textit{Handcent Next SMS messenger\footnote{https://play.google.com/store/apps/details?id=com.handcent.app.nextsms}} from the SMS category,
\textit{FOX Sports: Watch Live\footnote{https://play.google.com/store/apps/details?id=com.foxsports.android}} from the Sports News category, and  \textit{Weather \& Clock Widget\footnote{https://play.google.com/store/apps/details?id=com.devexpert.weather}} from the Weather app category. As shown in Table \ref{tab:rq2_suggest_sir}, the chosen Weather, SMS, and Sports News apps have 40, 140, and 36 release notes, respectively, with average word counts of 36, 30, and 23 per release. For each app, we apply steps 1 to 5 of \textit{LLM-Cure}'s Phase 2. We focus on the top three underperforming features that require the most attention from developers. For each feature, we identify three target complaints. For each user complaint, we generate suggestions to improve the app features. We obtain a total of 9 suggestions per feature, resulting in 27 suggestions per app. We then calculate \textit{SIR} for each feature.\\

\begin{table*}[t]
\centering
\scriptsize
\caption{Suggestions Implementation Rate (SIR) of \textit{LLM-Cure} Feature Improvement Suggestions on the selected three apps. Avr. denotes Average and Underperf. denotes Underperforming}
\label{tab:rq2_suggest_sir}
\begin{tabular}{llcccccc}
\toprule
\makecell{\textbf{App} \\ \textbf{category}} & \makecell{\textbf{App} \\ \textbf{name}} & \makecell{\textbf{\# of} \\ \textbf{releases}} & \makecell{\textbf{Avr. words} \\ \textbf{per release}} & \makecell{\textbf{Underperf.} \\ \textbf{features}} & \makecell{\textbf{\# of neg.} \\ \textbf{reviews}} & \makecell{\textbf{\# of pos.} \\ \textbf{reviews}} & \textbf{SIR} \\ 
\midrule
\multirow{3}{*}{Weather} & \multirow{3}{*}{Weather \& Clock Widget} & \multirow{3}{*}{44} &  \multirow{3}{*}{36} & Accuracy & 336 & 8,383 & 8/9 = 89\%\\
 & & & & Ease of Use & 159 & 4,313 & 5/9 = 56\% \\
 & & & & Performance & 157 & 1,951 & 4/9 =44\%\\
\midrule
\multirow{3}{*}{SMS} & \multirow{3}{*}{Handcent Next SMS Messenger} & \multirow{3}{*}{140}  &  \multirow{3}{*}{30} &  User Interface & 316 & 686 & 9/9 = 100\%\\
 & & & & Notifications & 297 & 359 & 6/9 = 67\% \\
 & & & & MMS Support & 229 & 162 & 7/9 = 78\%\\
\midrule
\multirow{3}{*}{Sports News} & \multirow{3}{*}{FOX Sports: Watch Live} & \multirow{3}{*}{36}  &  \multirow{3}{*}{23} & Performance & 27 & 1,947 & 8/9 = 89\%\\
 & & & & Sports Coverage & 26 & 5,640 & 4/9 = 44\% \\
 & & & & Notifications & 18 & 6,558 & 8/9 = 89\% \\
\midrule
\multicolumn{7}{l}{\textbf{Total SIR}} & \textbf{59/81 = 73\%} \\
\bottomrule
\end{tabular}
\end{table*}

\noindent\textbf{Results.} \textbf{\textit{LLM-Cure} achieves a promising \textit{SIR} of 59 out of 81 (i.e., 73\%), indicating the majority of the suggestionss from \textit{LLM-Cure} are implemented by the developers.} Table \ref{tab:rq2_suggest_sir} shows the SIR for each feature across the three apps. The results indicate that some features received higher \textit{SIRs} than others. For example, all the User Interface-related suggestions for \textit{"Handcent Next SMS messenger"} were implemented, while not all Notification-related suggestions for the same app were adopted. This variation can be attributed to different development priorities or challenges associated with certain feature enhancements. Furthermore, the release notes sometimes contained high-level descriptions of updates, such as "Performance improvement," which may not explicitly detail the changes but could reflect the overall enhancement suggested by \textit{LLM-Cure}.

\textbf{\textit{LLM-Cure} successfully leverages candidate positive user reviews from competitors to provide suggestions for feature improvements related to user complaints.} We illustrate this process through two case studies, randomly selecting underperforming features from the \textit{Weather \& Clock Widget} app. For the underperforming \textit{"Ease of Use"} feature, we consider the below user review that highlights an issue where the weather widget fails to update automatically, requiring manual refreshes:

\begin{center}
\begin{tcolorbox}[colback=black!5!white, colframe=black!75!black, boxrule=0.5mm, arc=4mm, left=2mm, right=2mm, top=3mm, bottom=3mm, width=0.7\textwidth] 
\begin{minipage}[t]{0.9\textwidth}
\begin{quote}
\leftskip=-20pt 
\rightskip=10pt 
\textit{``Widget won't update over time it keeps showing same weather and same TIME until I tap on it and open weather app.''}
\end{quote}
\vspace{2mm} 
\end{minipage}%
\begin{minipage}[t]{0.05\textwidth}
\vspace{-0.30em} 
\hfill
\starwithborder\unfilledstar\unfilledstar\unfilledstar\unfilledstar
\end{minipage}
\end{tcolorbox}
\end{center}

The target review identified two weaknesses in the widget: \textbf{(1) unreliable timekeeping} and \textbf{(2) an unfriendly design}. \textit{LLM-Cure} tackles these issues by proposing an automatic refresh functionality, an improved time synchronization, and the implementation of a refresh icon. These suggestions directly address the user's frustrations and aim to improve the widget's usability. Specifically \textit{LLM-Cure} proposes the below three suggestions:

\begin{center} 
\begin{tcolorbox}[colback=green!7!white, colframe=green!7!white,
  colbacktitle=green!50!black, title=LLM-Cure suggestions for the Ease of Use complaint,
  width=0.8\textwidth] 

  \vspace{4mm} 

  \textbf{Automatic Refresh}: Implement an automatic refresh feature for the widget that updates the weather and time data periodically, so users don't have to manually tap on the widget to see the current weather and time.\\

  \textbf{Improved Time Synchronization}: Enhance the clock functionality in the widget to ensure that it stays in sync with the device's internal clock, eliminating the need for manual time updates.\\

  \textbf{Clear Refresh Icon and Options}: Provide a clear refresh icon and add options for users to manually refresh the widget, change clock fonts, and customize the weather update intervals to their preference. This will give users more control over the widget's behavior and improve their overall experience.\\

\end{tcolorbox}
\end{center}

Subsequently, we cross-referenced these suggestions with the app's release notes and found that developers implemented recommendations provided by \textit{LLM-Cure} in future releases. For instance, in release \textit{6.1.0.1}, the functionalities \textit{"Added option to show forecast every 3 hours on the widget when you select hourly forecast"} and \textit{"Enabled digital font for clock and date"} were introduced, aligning closely with the suggestions aimed at enhancing the time synchronization of the widget. \\

Following the same approach outlined above, we select the below user complaint belonging to the \textit{Radar} underperforming feature:

\begin{center}
\begin{tcolorbox}[colback=black!5!white, colframe=black!75!black, boxrule=0.5mm, arc=4mm, left=2mm, right=2mm, top=3mm, bottom=3mm, width=0.7\textwidth] 
\begin{minipage}[t]{0.9\textwidth}
\vspace{2mm} 
\begin{quote}
\leftskip=-20pt 
\rightskip=10pt 
\leftskip=-20pt 
\rightskip=7pt 
\textit{"Needs a radar view for everyone's local area. Not having it takes away from the app."}
\end{quote}
\end{minipage}%
\begin{minipage}[t]{0.05\linewidth}
\vspace{-1.25em} 
\hfill
\starwithborder\starwithborder\unfilledstar\unfilledstar\unfilledstar
\end{minipage}
\end{tcolorbox}
\end{center}

The target review complains about the missing radar view for the local user area. The complaint highlights a shortcoming in the app's radar functionality on two fronts: \textbf{1) Radar location}, i.e., the user can't easily see weather patterns in their immediate surroundings and \textbf{2) Radar view perspectives}, i.e., the app lacks the flexibility to customize the radar display. \textit{LLM-Cure} addresses these issues by proposing customizable views, GPS integration, and real-time radar updates leveraging relative competitive user reviews. Table \ref{tab:candidate_reviews} illustrates a sample of the candidate reviews. 

Specifically, \textit{LLM-Cure} offer the below three suggestions:
\begin{center} 
\begin{tcolorbox}[colback=green!7!white, colframe=green!7!white,
  colbacktitle=green!50!black, title=LLM-Cure suggestions for the Radar complaint,
  width=0.8\textwidth]
  \textbf{Customizable Radar Views:} Some users have mentioned that they would like more control over the radar view, such as the ability to adjust the time frame or overlay different data layers. Adding customizable radar views could help users get the information they need for their specific location.\\

  \textbf{Integration with GPS:} Several users have praised radar features that integrate with GPS to automatically show their current location on the map. Adding this feature could help users quickly access radar information for their local area, even if they are not familiar with the region.\\

    \textbf{Real-time Radar Updates:} Many users have praised radar features that update in real-time, allowing them to track weather systems as they develop. Implementing real-time radar updates could help address the user's complaint about the app not having a radar view for their local area.
\end{tcolorbox}
\end{center}

\textit{LLM-Cure}'s suggestions directly target the weaknesses identified in the user complaint, demonstrating its ability to analyze user complaints and propose relevant improvements leveraging the reviews of competing apps. Furthermore, we analyzed the future release notes for the \textit{"Weather \& Clock Widget"} app after the user's complaint (\textit{release 5.9.1.3}). The improvements mentioned in future releases aligned with the LLM's suggestions. For instance, \textit{release 6.0.0.1} introduced a "weather radar service screen," directly addressing the lack of a local radar view. Additionally, \textit{release 6.0.1.2} offered the \textit{"option to set radar default layer"}, which aligns with the suggestion for customizable views. 

\begin{table}[h]
\centering
\caption{A sample of five reviews regarding the \textit{Radar} feature from competitors of the \textit{"Weather \& Clock Widget"} app}
\label{tab:candidate_reviews}
\begin{tabular}{>{\raggedright\arraybackslash\textit{}}p{\linewidth}} 
\toprule
\textbf{Competitors' user reviews} \\
\midrule
\rowcolor{gray!20} 
\textit{"I like that the radar is on the first page. Lots of times I just
want to see where the rain/snow/storm is and how close it is. I
don’t have to go through a bunch of screens to get there."}\\
\textit{"I love that it puts my current location on the radar maps. Great interface."}\\
\rowcolor{gray!20} 
\textit{"Radar update with gps location is very accurate. It helps me in my road trip planning avoiding snow hazards. Loved it. It has variety of radar scans temperature etc."}\\
\textit{"I like the live radar. Wish it was a longer time Frame though."} \\
\rowcolor{gray!20} 
\textit{"Exactly what I need. Easy to find radar. Easy to view by location or other areas."}\\
\bottomrule
\end{tabular}
\end{table}

\textbf{\textit{LLM-Cure's} suggestions often align with functionalities later implemented by the apps documented in the release notes.} Table \ref{tab:suggest_release_notes_matches} presents a random sample of \textit{LLM-Cure} suggestions alongside corresponding release notes selected from 81 of the total suggestions, demonstrating the alignment between the feature improvement suggestions and the actual implementations. These results provide encouraging evidence that \textit{LLM-Cure} can effectively analyze competitors' user reviews and generate suggestions for feature improvements.

\begin{table}[H] 
\centering
\caption{Sample of LLM-Cure's suggestions and corresponding release notes}
\label{tab:suggest_release_notes_matches}
\begin{tabular}{p{0.5\linewidth}p{0.45\linewidth}} 
\toprule
\textbf{LLM-Cure's Suggestions} & \textbf{Release Notes}\\ 
\midrule
\rowcolor{gray!20}
Add a customizable notification sound: Allowing users to set their preferred notification sound can significantly improve their experience & Fixed personalized sound notification\\
\midrule
\rowcolor{white} 
Improve the quality of MMS messages: The app should allow users to send high-quality images and videos through MMS & High quality picture compression when MMS\\
\midrule
\rowcolor{gray!20}
Reliable MMS delivery: Improve the MMS delivery system to ensure that messages are sent and received reliably & Improved send/receive MMS known issues\\
\midrule
\rowcolor{white} 
Improve international sports coverage: Some users have requested better coverage of international sports, such as rugby and cricket & Expanded golf coverage including schedule, leaderboards, scorecards, tee times, and rankings\\
\midrule
\rowcolor{gray!20}
Add a dedicated tab for NHRA events: To address the user complaint, the app could add a specific tab for NHRA events, making it easier for users to find and access coverage of these events & Enhanced home screen with all your favorite team scores, news, and live streams in one place\\
\bottomrule
\end{tabular}
\end{table}

\begin{Summary}{}{firstsummary}
\textit{LLM-Cure} successfully leverages candidate positive reviews of competitors to generate feature improvement suggestions for user complaints. \textit{LLM-Cure} achieves a promising average of 73\% of Suggestions Implementation Rate (SIR), demonstrating its potential for competitive feature enhancement.
\end{Summary}

\section{Threats to Validity}
\label{sec:threats}
\noindent\textbf{Threats to construct validity} relate to a possible error in the data preparation. In \textit{LLM-Cure}, we adopt a batch-and-match methodology to accommodate a scalable LLM prompting by employing a subset of reviews to extract features. We evaluate the validity of our approach by testing it across various thresholds and validating against ground truth data, ensuring robustness in representation. To reduce any bias that may be introduced by the order of reviews, we shuffle user reviews before splitting them to ensure a representative sample of user reviews.

In Phase 2, i.e., the Suggestion Generation with Competitor
Reviews phase, we suggest addressing the most underperforming features, i.e., having the highest number of negative reviews in the target app. We acknowledge that developers may employ different prioritization (e.g., severity-based, effort-based) techniques in the real world. However, our methodology remains valid. Our approach is anchored in the actual reviews rather than the selection approach, thus ensuring the reliability of our results. Developers have the flexibility to prioritize and address complaints from any features, whether they are functional or non-functional.\\

\noindent\textbf{Threats to internal validity} relate to the concerns that might come from the internal methods used in our study. One threat may stem from the selection and design of the prompt templates. To address this potential threat, we explore different prompts. When provided with the same shot examples, we observe that the prompt template does not lead to different classification results. All the used prompts are available in the replication package.\\
 
\noindent\textbf{Threats to external validity} concern the ability to generalize the results. One threat concerns the choice of the LLM utilized in our approach, \textit{LLM-Cure}. We acknowledge that each LLM possesses a distinct architecture, potentially leading to variations in results. However, we opt for Mistral for several reasons. As detailed in Section \ref{sub:implementation}, Mistral outperforms other models across various benchmarks. Furthermore, Mistral is openly available for researchers, fostering the replication of our work and ensuring transparency in the evaluation process. Despite these considerations, it's important to recognize that the choice of LLM remains a potential source of variability in our findings. 
Another threat to external validity pertains to the selection of categories and datasets. However, similar to prior research\cite{assi2021featcompare}, we mitigate this concern by evaluating the validity of our approach across five distinct app categories sourced from ground truth data. This approach aims to minimize the influence of app selection bias. Additionally, our method is platform-agnostic, offering applicability to any mobile app, provided it contains user reviews. This broad applicability enhances the generalizability of our findings beyond specific categories or datasets.

\section{Related Work}
\label{sec:related_work}
\noindent\textbf{Feature Enhancement.} Prior approaches have explored automatic feature extraction from user reviews for feature enhancement \cite{9920051, WANG2022118095, https://doi.org/10.1002/smr.2257, LIU2021129}. For example, Scalabrino et al. \cite{9920051} introduce CLAP, a web application facilitating mobile app release planning by analyzing user reviews. CLAP categorizes reviews from the target app, prioritizing user concerns to be addressed. Wang et al. \cite{WANG2022118095} present UISMiner, which supports UI-related feature enhancement by mining user review suggestions about UI. Gao et al. \cite{https://doi.org/10.1002/smr.2257} propose a method for analyzing user reviews to extract requirements and update app goal models, including feature improvements and additions. However, these contributions focus solely on individual target apps, lacking consideration of competing apps. Liu et al. \cite{LIU2021129} present an approach considering market trends, guiding developers on feature update strategies by comparing features of similar apps. While Liu et al.'s work includes a competitive analysis aspect, it primarily suggests which features to update rather than offering suggestions for improvement.

Similar to the aforementioned contributions, our aim is feature enhancement. However, our work distinguishes itself in two key aspects: (1) we aim to guide developers towards feature enhancements automatically, irrespective of the feature categories; and (2) we offer a competitive landscape for feature improvement by harnessing the competitors' user reviews and the power of LLMs. This allows us to not only suggest enhancements based on user complaints but also provide insights into how these features compare with those of competing apps.\\


\noindent\textbf{Mobile Apps Competitor Feature Analysis.} Researchers have proposed methods for extracting detailed, fine-grained features from user reviews and comparing them across app competitors. Shah et al. \cite{ShahSP19} introduce REVSUM, a competitor analysis tool that evaluates sentiment, bug reports, and feature requests between a target app and its competitors. Dalpiaz and Parente \cite{DBLP:conf/refsq/DalpiazP19} introduce RE-SWOT, which constructs a Strength-Weakness-Opportunity-Threat (SWOT) matrix for competitor analysis. Another tool by Shah et al. \cite{DBLP:conf/sigsoft/ShahSP16} compares two competing apps based on the selected features and analyzes competitor sentiment. However, relying on fine-grained features for competitor analysis is impractical due to the large number of features that can be extracted \cite{assi2021featcompare, DBLP:conf/refsq/DalpiazP19, DBLP:journals/corr/abs-1810-05187}. In response, Featcompare \cite{assi2021featcompare} proposes an automatic approach to mine high-level features (i.e., semantic clusters of fine-grained features) from competing apps and compares user sentiment across these features. However, none of these approaches offer developers suggestions for improving features. Wang et al. \cite{WANG2022111435} introduce a UI-focused feature recommender tool for enhancing app competitiveness by recommending missing features based on the analysis of UI pages from similar apps. While Wang et al.'s work recommends missing features from a target app, it is only limited to the UI features.\\

\noindent\textbf{LLMs for Mobile Apps.} De Lima et al. \cite{Lima2023} propose a method utilizing LLMs to autonomously identify risk factors from app reviews and prioritize them to anticipate and mitigate risks. Roumeliotis et al. \cite{ROUMELIOTIS2024100056} conduct an evaluation study comparing the effectiveness of LLMs like Llama and GPT 3.5 in predicting sentiment analysis related to e-commerce. Similarly, Zhang et al. \cite{zhang2023sentiment} assess the performance of three open-source LLMs in zero-shot and few-shot settings for predicting sentiment in user reviews. Xu et al. \cite{xu2023limits} design a prompt instructing ChatGPT to extract aspect-category-opinion-sentiment quadruples from text. Wei et al. \cite{10356483} propose Mini-BAR, a tool integrating LLMs for zero-shot mining of bilingual user reviews in English and French. Dos Santos et al. \cite{10.1145/3638067.3638081} analyze accessibility reviews using LLMs. While the above work leverages user reviews, Huang et al. \cite{10356483} introduce CrashTranslator, which automatically reproduces mobile app crashes from stack traces guided by LLMs to predict the exploration steps for triggering the crash. Liu et al. \cite{Liu_2024} propose InputBlaster, leveraging LLMs to generate unusual text inputs for mobile app crash detection. In our approach, we harness LLMs to provide suggestions for feature improvement, complementing the aforementioned contributions to support mobile app analysis.

\section{Conclusion}
\label{sec:conclusion}
\noindent In this work, we introduce \textit{LLM-Cure}, a novel LLM-based approach that conducts Competitor User Review Analysis for Feature Enhancement. \textit{LLM-Cure} generates automated suggestions for app feature improvements by leveraging user reviews from competitor apps to enhance user experience and maintain competitiveness. We evaluate \textit{LLM-Cure} on 1,056,739 reviews of 70 popular Android apps. \textit{LLM-Cure} achieves high performance in extracting and assigning features to user reviews, outperforming baseline methods significantly by up to 13\% in F1-score, up to 16\% in recall and up to 11\% in
precision. \textit{LLM-Cure} achieves a promising suggestions implementation rate of 58 out of 81. By combining user feedback with competitor user review analysis, \textit{LLM-Cure} empowers developers to make informed decisions, fostering a more competitive landscape.

In the future, we aim to explore agent orchestration between LLMs to streamline the feature analysis process and leverage the strengths of different language models. Additionally, we aim to expand \textit{LLM-Cure}'s capabilities to prioritize suggestions based on factors, such as popularity and potential user impact.\\

\noindent \textbf{Data Availability.} We provide a replication package including the data and scripts to replicate the analyses at \url{https://github.com/repl-pack/LLM-Cure}. 


\bibliographystyle{ACM-Reference-Format}
\bibliography{sample-base}


\begin{thebibliography}{64}


\ifx \showCODEN    \undefined \def \showCODEN     #1{\unskip}     \fi
\ifx \showDOI      \undefined \def \showDOI       #1{#1}\fi
\ifx \showISBNx    \undefined \def \showISBNx     #1{\unskip}     \fi
\ifx \showISBNxiii \undefined \def \showISBNxiii  #1{\unskip}     \fi
\ifx \showISSN     \undefined \def \showISSN      #1{\unskip}     \fi
\ifx \showLCCN     \undefined \def \showLCCN      #1{\unskip}     \fi
\ifx \shownote     \undefined \def \shownote      #1{#1}          \fi
\ifx \showarticletitle \undefined \def \showarticletitle #1{#1}   \fi
\ifx \showURL      \undefined \def \showURL       {\relax}        \fi
\providecommand\bibfield[2]{#2}
\providecommand\bibinfo[2]{#2}
\providecommand\natexlab[1]{#1}
\providecommand\showeprint[2][]{arXiv:#2}

\bibitem[AI(2024)]%
        {AI_2024}
\bibfield{author}{\bibinfo{person}{Mistral AI}.} \bibinfo{year}{2024}\natexlab{}.
\newblock \bibinfo{title}{Mixtral of Experts}.
\newblock
\newblock
\urldef\tempurl%
\url{https://mistral.ai/news/mixtral-of-experts/}
\showURL{%
\tempurl}


\bibitem[{AlSubaihin} et~al\mbox{.}(2019)]%
        {8606261}
\bibfield{author}{\bibinfo{person}{A. {AlSubaihin}}, \bibinfo{person}{F. {Sarro}}, \bibinfo{person}{S. {Black}}, \bibinfo{person}{L. {Capra}}, {and} \bibinfo{person}{M. {Harman}}.} \bibinfo{year}{2019}\natexlab{}.
\newblock \showarticletitle{App Store Effects on Software Engineering Practices}.
\newblock \bibinfo{journal}{\emph{IEEE Transactions on Software Engineering}} (\bibinfo{year}{2019}), \bibinfo{pages}{1--1}.
\newblock
\urldef\tempurl%
\url{https://doi.org/10.1109/TSE.2019.2891715}
\showDOI{\tempurl}


\bibitem[Aralikatte et~al\mbox{.}(2018)]%
        {AralikatteSGM18}
\bibfield{author}{\bibinfo{person}{Rahul Aralikatte}, \bibinfo{person}{Giriprasad Sridhara}, \bibinfo{person}{Neelamadhav Gantayat}, {and} \bibinfo{person}{Senthil Mani}.} \bibinfo{year}{2018}\natexlab{}.
\newblock \showarticletitle{Fault in your stars: an analysis of Android app reviews}. In \bibinfo{booktitle}{\emph{Proceedings of the {ACM} India Joint International Conference on Data Science and Management of Data, {COMAD/CODS} 2018, Goa, India, January 11-13, 2018}}, \bibfield{editor}{\bibinfo{person}{Sayan Ranu}, \bibinfo{person}{Niloy Ganguly}, \bibinfo{person}{Raghu Ramakrishnan}, \bibinfo{person}{Sunita Sarawagi}, {and} \bibinfo{person}{Shourya Roy}} (Eds.). \bibinfo{publisher}{{ACM}}, \bibinfo{pages}{57--66}.
\newblock
\urldef\tempurl%
\url{https://doi.org/10.1145/3152494.3152500}
\showDOI{\tempurl}


\bibitem[Arambepola et~al\mbox{.}(2024)]%
        {10499727}
\bibfield{author}{\bibinfo{person}{Nimasha Arambepola}, \bibinfo{person}{Lankeshwara Munasinghe}, {and} \bibinfo{person}{Nalin Warnajith}.} \bibinfo{year}{2024}\natexlab{}.
\newblock \showarticletitle{Factors Influencing Mobile App User Experience: An Analysis of Education App User Reviews}. In \bibinfo{booktitle}{\emph{2024 4th International Conference on Advanced Research in Computing (ICARC)}}. \bibinfo{pages}{223--228}.
\newblock
\urldef\tempurl%
\url{https://doi.org/10.1109/ICARC61713.2024.10499727}
\showDOI{\tempurl}


\bibitem[Assi et~al\mbox{.}(2023)]%
        {10.1145/3593802}
\bibfield{author}{\bibinfo{person}{Maram Assi}, \bibinfo{person}{Safwat Hassan}, \bibinfo{person}{Stefanos Georgiou}, {and} \bibinfo{person}{Ying Zou}.} \bibinfo{year}{2023}\natexlab{}.
\newblock \showarticletitle{Predicting the Change Impact of Resolving Defects by Leveraging the Topics of Issue Reports in Open Source Software Systems}.
\newblock \bibinfo{journal}{\emph{ACM Trans. Softw. Eng. Methodol.}} \bibinfo{volume}{32}, \bibinfo{number}{6}, Article \bibinfo{articleno}{141} (\bibinfo{date}{sep} \bibinfo{year}{2023}), \bibinfo{numpages}{34}~pages.
\newblock
\showISSN{1049-331X}
\urldef\tempurl%
\url{https://doi.org/10.1145/3593802}
\showDOI{\tempurl}


\bibitem[Assi et~al\mbox{.}(2021)]%
        {assi2021featcompare}
\bibfield{author}{\bibinfo{person}{Maram Assi}, \bibinfo{person}{Safwat Hassan}, \bibinfo{person}{Yuan Tian}, {and} \bibinfo{person}{Ying Zou}.} \bibinfo{year}{2021}\natexlab{}.
\newblock \showarticletitle{FeatCompare: Feature comparison for competing mobile apps leveraging user reviews}.
\newblock \bibinfo{journal}{\emph{Empirical Software Engineering}} \bibinfo{volume}{26}, \bibinfo{number}{5} (\bibinfo{year}{2021}).
\newblock


\bibitem[Bengio et~al\mbox{.}(2003)]%
        {Bengio2003ANP}
\bibfield{author}{\bibinfo{person}{Yoshua Bengio}, \bibinfo{person}{R{\'e}jean Ducharme}, \bibinfo{person}{Pascal Vincent}, {and} \bibinfo{person}{Christian Janvin}.} \bibinfo{year}{2003}\natexlab{}.
\newblock \showarticletitle{A Neural Probabilistic Language Model}.
\newblock \bibinfo{journal}{\emph{J. Mach. Learn. Res.}}  \bibinfo{volume}{3} (\bibinfo{year}{2003}), \bibinfo{pages}{1137--1155}.
\newblock
\urldef\tempurl%
\url{https://api.semanticscholar.org/CorpusID:221275765}
\showURL{%
\tempurl}


\bibitem[Brown et~al\mbox{.}(2020)]%
        {brown2020language}
\bibfield{author}{\bibinfo{person}{Tom~B. Brown}, \bibinfo{person}{Benjamin Mann}, \bibinfo{person}{Nick Ryder}, \bibinfo{person}{Melanie Subbiah}, \bibinfo{person}{Jared Kaplan}, \bibinfo{person}{Prafulla Dhariwal}, \bibinfo{person}{Arvind Neelakantan}, \bibinfo{person}{Pranav Shyam}, \bibinfo{person}{Girish Sastry}, \bibinfo{person}{Amanda Askell}, \bibinfo{person}{Sandhini Agarwal}, \bibinfo{person}{Ariel Herbert-Voss}, \bibinfo{person}{Gretchen Krueger}, \bibinfo{person}{Tom Henighan}, \bibinfo{person}{Rewon Child}, \bibinfo{person}{Aditya Ramesh}, \bibinfo{person}{Daniel~M. Ziegler}, \bibinfo{person}{Jeffrey Wu}, \bibinfo{person}{Clemens Winter}, \bibinfo{person}{Christopher Hesse}, \bibinfo{person}{Mark Chen}, \bibinfo{person}{Eric Sigler}, \bibinfo{person}{Mateusz Litwin}, \bibinfo{person}{Scott Gray}, \bibinfo{person}{Benjamin Chess}, \bibinfo{person}{Jack Clark}, \bibinfo{person}{Christopher Berner}, \bibinfo{person}{Sam McCandlish}, \bibinfo{person}{Alec Radford}, \bibinfo{person}{Ilya Sutskever},
  {and} \bibinfo{person}{Dario Amodei}.} \bibinfo{year}{2020}\natexlab{}.
\newblock \bibinfo{title}{Language Models are Few-Shot Learners}.
\newblock
\newblock
\showeprint[arxiv]{2005.14165}~[cs.CL]


\bibitem[by~Laura~Ceci and 4(2024)]%
        {Laura}
\bibfield{author}{\bibinfo{person}{Published by Laura~Ceci} {and} \bibinfo{person}{Mar 4}.} \bibinfo{year}{2024}\natexlab{}.
\newblock \bibinfo{title}{Messaging apps: Most popular by global downloads 2024}.
\newblock
\newblock
\urldef\tempurl%
\url{https://www.statista.com/statistics/1263360/most-popular-messenger-apps-worldwide-by-monthly-downloads/}
\showURL{%
\tempurl}


\bibitem[Ceci(2024)]%
        {Ceci}
\bibfield{author}{\bibinfo{person}{Laura Ceci}.} \bibinfo{year}{2024}\natexlab{}.
\newblock \bibinfo{title}{Number of mobile app downloads worldwide from 2016 to 2023}.
\newblock \bibinfo{howpublished}{\url{https://www.https://www.statista.com/statistics/271644/worldwide-free-and-paid-mobile-app-store-downloads/}}.
\newblock
\newblock
\shownote{(Last accessed May 2024)}.


\bibitem[Chen et~al\mbox{.}(2014)]%
        {DBLP:conf/icse/ChenLHXZ14}
\bibfield{author}{\bibinfo{person}{Ning Chen}, \bibinfo{person}{Jialiu Lin}, \bibinfo{person}{Steven C.~H. Hoi}, \bibinfo{person}{Xiaokui Xiao}, {and} \bibinfo{person}{Boshen Zhang}.} \bibinfo{year}{2014}\natexlab{}.
\newblock \showarticletitle{{AR-miner}: mining informative reviews for developers from mobile app marketplace}. In \bibinfo{booktitle}{\emph{Proceedings of the 36th International Conference on Software Engineering}} \emph{(\bibinfo{series}{ICSE '14})}. \bibinfo{pages}{767--778}.
\newblock


\bibitem[Cohen(1960)]%
        {Cohen}
\bibfield{author}{\bibinfo{person}{Jacob Cohen}.} \bibinfo{year}{1960}\natexlab{}.
\newblock \showarticletitle{A Coefficient of Agreement for Nominal Scales}.
\newblock \bibinfo{journal}{\emph{Educational and Psychological Measurement}} \bibinfo{volume}{20}, \bibinfo{number}{1} (\bibinfo{year}{1960}), \bibinfo{pages}{37--46}.
\newblock


\bibitem[Dalpiaz and Parente(2019)]%
        {DBLP:conf/refsq/DalpiazP19}
\bibfield{author}{\bibinfo{person}{Fabiano Dalpiaz} {and} \bibinfo{person}{Micaela Parente}.} \bibinfo{year}{2019}\natexlab{}.
\newblock \showarticletitle{{RE-SWOT:} From User Feedback to Requirements via Competitor Analysis}. In \bibinfo{booktitle}{\emph{Proceedings of the 25th International Working Conference on Requirements Engineering: Foundation for Software Quality}} \emph{(\bibinfo{series}{REFSQ '19}, Vol.~\bibinfo{volume}{11412})}. \bibinfo{pages}{55--70}.
\newblock


\bibitem[DeFilippis et~al\mbox{.}(2022)]%
        {defilippis2022impact}
\bibfield{author}{\bibinfo{person}{Evan DeFilippis}, \bibinfo{person}{Stephen~Michael Impink}, \bibinfo{person}{Madison Singell}, \bibinfo{person}{Jeffrey~T Polzer}, {and} \bibinfo{person}{Raffaella Sadun}.} \bibinfo{year}{2022}\natexlab{}.
\newblock \showarticletitle{The impact of COVID-19 on digital communication patterns}.
\newblock \bibinfo{journal}{\emph{Humanities and Social Sciences Communications}} \bibinfo{volume}{9}, \bibinfo{number}{1} (\bibinfo{year}{2022}).
\newblock


\bibitem[Devine et~al\mbox{.}(2022)]%
        {9920051}
\bibfield{author}{\bibinfo{person}{Peter Devine}, \bibinfo{person}{James Tizard}, \bibinfo{person}{Hechen Wang}, \bibinfo{person}{Yun~Sing Koh}, {and} \bibinfo{person}{Kelly Blincoe}.} \bibinfo{year}{2022}\natexlab{}.
\newblock \showarticletitle{What’s Inside a Cluster of Software User Feedback: A Study of Characterisation Methods}. In \bibinfo{booktitle}{\emph{2022 IEEE 30th International Requirements Engineering Conference (RE)}}. \bibinfo{pages}{189--200}.
\newblock
\urldef\tempurl%
\url{https://doi.org/10.1109/RE54965.2022.00023}
\showDOI{\tempurl}


\bibitem[Dong et~al\mbox{.}(2023b)]%
        {dong2023survey}
\bibfield{author}{\bibinfo{person}{Qingxiu Dong}, \bibinfo{person}{Lei Li}, \bibinfo{person}{Damai Dai}, \bibinfo{person}{Ce Zheng}, \bibinfo{person}{Zhiyong Wu}, \bibinfo{person}{Baobao Chang}, \bibinfo{person}{Xu Sun}, \bibinfo{person}{Jingjing Xu}, \bibinfo{person}{Lei Li}, {and} \bibinfo{person}{Zhifang Sui}.} \bibinfo{year}{2023}\natexlab{b}.
\newblock \bibinfo{title}{A Survey on In-context Learning}.
\newblock
\newblock
\showeprint[arxiv]{2301.00234}~[cs.CL]


\bibitem[Dong et~al\mbox{.}(2023a)]%
        {dong2023selfcollaboration}
\bibfield{author}{\bibinfo{person}{Yihong Dong}, \bibinfo{person}{Xue Jiang}, \bibinfo{person}{Zhi Jin}, {and} \bibinfo{person}{Ge Li}.} \bibinfo{year}{2023}\natexlab{a}.
\newblock \bibinfo{title}{Self-collaboration Code Generation via ChatGPT}.
\newblock
\newblock
\showeprint[arxiv]{2304.07590}~[cs.SE]


\bibitem[Dos~Santos et~al\mbox{.}(2024)]%
        {10.1145/3638067.3638081}
\bibfield{author}{\bibinfo{person}{Paulo S\'{e}rgio~Henrique Dos~Santos}, \bibinfo{person}{Alberto Dumont~Alves Oliveira}, \bibinfo{person}{Thais Bonjorni~Nobre De~Jesus}, \bibinfo{person}{Wajdi Aljedaani}, {and} \bibinfo{person}{Marcelo~Medeiros Eler}.} \bibinfo{year}{2024}\natexlab{}.
\newblock \showarticletitle{Evolution may come with a price: analyzing user reviews to understand the impact of updates on mobile apps accessibility}. In \bibinfo{booktitle}{\emph{Proceedings of the XXII Brazilian Symposium on Human Factors in Computing Systems}} (<conf-loc>, <city>Macei\'{o}</city>, <country>Brazil</country>, </conf-loc>) \emph{(\bibinfo{series}{IHC '23})}. \bibinfo{publisher}{Association for Computing Machinery}, \bibinfo{address}{New York, NY, USA}, Article \bibinfo{articleno}{52}, \bibinfo{numpages}{11}~pages.
\newblock
\showISBNx{9798400717154}
\urldef\tempurl%
\url{https://doi.org/10.1145/3638067.3638081}
\showDOI{\tempurl}


\bibitem[Fan et~al\mbox{.}(2022)]%
        {Fan2022ImprovingAG}
\bibfield{author}{\bibinfo{person}{Zhiyu Fan}, \bibinfo{person}{Xiang Gao}, \bibinfo{person}{Abhik Roychoudhury}, {and} \bibinfo{person}{Shin~Hwei Tan}.} \bibinfo{year}{2022}\natexlab{}.
\newblock \showarticletitle{Improving automatically generated code from Codex via Automated Program Repair}.
\newblock \bibinfo{journal}{\emph{ArXiv}}  \bibinfo{volume}{abs/2205.10583} (\bibinfo{year}{2022}).
\newblock
\urldef\tempurl%
\url{https://api.semanticscholar.org/CorpusID:248986410}
\showURL{%
\tempurl}


\bibitem[Fu et~al\mbox{.}(2013a)]%
        {fu2013people}
\bibfield{author}{\bibinfo{person}{Bin Fu}, \bibinfo{person}{Jialiu Lin}, \bibinfo{person}{Lei Li}, \bibinfo{person}{Christos Faloutsos}, \bibinfo{person}{Jason Hong}, {and} \bibinfo{person}{Norman Sadeh}.} \bibinfo{year}{2013}\natexlab{a}.
\newblock \showarticletitle{Why people hate your app: Making sense of user feedback in a mobile app store}. In \bibinfo{booktitle}{\emph{Proceedings of the 19th ACM SIGKDD International Conference on Knowledge Discovery and Data Mining}} \emph{(\bibinfo{series}{KDD '13})}. \bibinfo{pages}{1276--1284}.
\newblock


\bibitem[Fu et~al\mbox{.}(2013b)]%
        {10.1145/2487575.2488202}
\bibfield{author}{\bibinfo{person}{Bin Fu}, \bibinfo{person}{Jialiu Lin}, \bibinfo{person}{Lei Li}, \bibinfo{person}{Christos Faloutsos}, \bibinfo{person}{Jason Hong}, {and} \bibinfo{person}{Norman Sadeh}.} \bibinfo{year}{2013}\natexlab{b}.
\newblock \showarticletitle{Why people hate your app: making sense of user feedback in a mobile app store}. In \bibinfo{booktitle}{\emph{Proceedings of the 19th ACM SIGKDD International Conference on Knowledge Discovery and Data Mining}} (Chicago, Illinois, USA) \emph{(\bibinfo{series}{KDD '13})}. \bibinfo{publisher}{Association for Computing Machinery}, \bibinfo{address}{New York, NY, USA}, \bibinfo{pages}{1276–1284}.
\newblock
\showISBNx{9781450321747}
\urldef\tempurl%
\url{https://doi.org/10.1145/2487575.2488202}
\showDOI{\tempurl}


\bibitem[Gao et~al\mbox{.}(2023)]%
        {9952173}
\bibfield{author}{\bibinfo{person}{Cuiyun Gao}, \bibinfo{person}{Yaoxian Li}, \bibinfo{person}{Shuhan Qi}, \bibinfo{person}{Yang Liu}, \bibinfo{person}{Xuan Wang}, \bibinfo{person}{Zibin Zheng}, {and} \bibinfo{person}{Qing Liao}.} \bibinfo{year}{2023}\natexlab{}.
\newblock \showarticletitle{Listening to Users' Voice: Automatic Summarization of Helpful App Reviews}.
\newblock \bibinfo{journal}{\emph{IEEE Transactions on Reliability}} \bibinfo{volume}{72}, \bibinfo{number}{4} (\bibinfo{year}{2023}), \bibinfo{pages}{1619--1631}.
\newblock
\urldef\tempurl%
\url{https://doi.org/10.1109/TR.2022.3217566}
\showDOI{\tempurl}


\bibitem[Gao et~al\mbox{.}(2019)]%
        {8804432}
\bibfield{author}{\bibinfo{person}{Cuiyun Gao}, \bibinfo{person}{Wujie Zheng}, \bibinfo{person}{Yuetang Deng}, \bibinfo{person}{David Lo}, \bibinfo{person}{Jichuan Zeng}, \bibinfo{person}{Michael~R. Lyu}, {and} \bibinfo{person}{Irwin King}.} \bibinfo{year}{2019}\natexlab{}.
\newblock \showarticletitle{Emerging App Issue Identification from User Feedback: Experience on WeChat}. In \bibinfo{booktitle}{\emph{2019 IEEE/ACM 41st International Conference on Software Engineering: Software Engineering in Practice (ICSE-SEIP)}}. \bibinfo{pages}{279--288}.
\newblock
\urldef\tempurl%
\url{https://doi.org/10.1109/ICSE-SEIP.2019.00040}
\showDOI{\tempurl}


\bibitem[Gao et~al\mbox{.}(2020)]%
        {https://doi.org/10.1002/smr.2257}
\bibfield{author}{\bibinfo{person}{Shanquan Gao}, \bibinfo{person}{Lei Liu}, \bibinfo{person}{Yuzhou Liu}, \bibinfo{person}{Huaxiao Liu}, {and} \bibinfo{person}{Yihui Wang}.} \bibinfo{year}{2020}\natexlab{}.
\newblock \showarticletitle{Updating the goal model with user reviews for the evolution of an app}.
\newblock \bibinfo{journal}{\emph{Journal of Software: Evolution and Process}} \bibinfo{volume}{32}, \bibinfo{number}{8} (\bibinfo{year}{2020}), \bibinfo{pages}{e2257}.
\newblock
\urldef\tempurl%
\url{https://doi.org/10.1002/smr.2257}
\showDOI{\tempurl}
\showeprint{https://onlinelibrary.wiley.com/doi/pdf/10.1002/smr.2257}
\newblock
\shownote{e2257 JSME-19-0105.R2}.


\bibitem[He et~al\mbox{.}(2017)]%
        {he-etal-2017-unsupervised}
\bibfield{author}{\bibinfo{person}{Ruidan He}, \bibinfo{person}{Wee~Sun Lee}, \bibinfo{person}{Hwee~Tou Ng}, {and} \bibinfo{person}{Daniel Dahlmeier}.} \bibinfo{year}{2017}\natexlab{}.
\newblock \showarticletitle{An Unsupervised Neural Attention Model for Aspect Extraction}. In \bibinfo{booktitle}{\emph{Proceedings of the 55th Annual Meeting of the Association for Computational Linguistics (Volume 1: Long Papers)}} \emph{(\bibinfo{series}{ACL '17})}. \bibinfo{pages}{388--397}.
\newblock


\bibitem[Hou et~al\mbox{.}(2024)]%
        {hou2024large}
\bibfield{author}{\bibinfo{person}{Xinyi Hou}, \bibinfo{person}{Yanjie Zhao}, \bibinfo{person}{Yue Liu}, \bibinfo{person}{Zhou Yang}, \bibinfo{person}{Kailong Wang}, \bibinfo{person}{Li Li}, \bibinfo{person}{Xiapu Luo}, \bibinfo{person}{David Lo}, \bibinfo{person}{John Grundy}, {and} \bibinfo{person}{Haoyu Wang}.} \bibinfo{year}{2024}\natexlab{}.
\newblock \bibinfo{title}{Large Language Models for Software Engineering: A Systematic Literature Review}.
\newblock
\newblock
\showeprint[arxiv]{2308.10620}~[cs.SE]


\bibitem[Hu et~al\mbox{.}(2021)]%
        {hu2021lora}
\bibfield{author}{\bibinfo{person}{Edward~J. Hu}, \bibinfo{person}{Yelong Shen}, \bibinfo{person}{Phillip Wallis}, \bibinfo{person}{Zeyuan Allen-Zhu}, \bibinfo{person}{Yuanzhi Li}, \bibinfo{person}{Shean Wang}, \bibinfo{person}{Lu Wang}, {and} \bibinfo{person}{Weizhu Chen}.} \bibinfo{year}{2021}\natexlab{}.
\newblock \bibinfo{title}{LoRA: Low-Rank Adaptation of Large Language Models}.
\newblock
\newblock
\showeprint[arxiv]{2106.09685}~[cs.CL]


\bibitem[Huang et~al\mbox{.}(2023)]%
        {huang2023survey}
\bibfield{author}{\bibinfo{person}{Lei Huang}, \bibinfo{person}{Weijiang Yu}, \bibinfo{person}{Weitao Ma}, \bibinfo{person}{Weihong Zhong}, \bibinfo{person}{Zhangyin Feng}, \bibinfo{person}{Haotian Wang}, \bibinfo{person}{Qianglong Chen}, \bibinfo{person}{Weihua Peng}, \bibinfo{person}{Xiaocheng Feng}, \bibinfo{person}{Bing Qin}, {and} \bibinfo{person}{Ting Liu}.} \bibinfo{year}{2023}\natexlab{}.
\newblock \bibinfo{title}{A Survey on Hallucination in Large Language Models: Principles, Taxonomy, Challenges, and Open Questions}.
\newblock
\newblock
\showeprint[arxiv]{2311.05232}~[cs.CL]


\bibitem[Iacob and Harrison(2013)]%
        {DBLP:conf/msr/IacobH13}
\bibfield{author}{\bibinfo{person}{Claudia Iacob} {and} \bibinfo{person}{Rachel Harrison}.} \bibinfo{year}{2013}\natexlab{}.
\newblock \showarticletitle{Retrieving and analyzing mobile apps feature requests from online reviews}. In \bibinfo{booktitle}{\emph{Proceedings of the 10th Working Conference on Mining Software Repositories}} \emph{(\bibinfo{series}{MSR '13})}. \bibinfo{pages}{41--44}.
\newblock


\bibitem[{Johann} et~al\mbox{.}(2017)]%
        {8048887}
\bibfield{author}{\bibinfo{person}{T. {Johann}}, \bibinfo{person}{C. {Stanik}}, \bibinfo{person}{A.~M.~A. {B.}}, {and} \bibinfo{person}{W. {Maalej}}.} \bibinfo{year}{2017}\natexlab{}.
\newblock \showarticletitle{{SAFE}: A Simple Approach for Feature Extraction from App Descriptions and App Reviews}. In \bibinfo{booktitle}{\emph{Proceedings of the 25th International Requirements Engineering Conference}} \emph{(\bibinfo{series}{RE '17})}. \bibinfo{pages}{21--30}.
\newblock


\bibitem[Li et~al\mbox{.}(2017)]%
        {Li}
\bibfield{author}{\bibinfo{person}{Yuanchun Li}, \bibinfo{person}{Baoxiong Jia}, \bibinfo{person}{Yao Guo}, {and} \bibinfo{person}{Xiangqun Chen}.} \bibinfo{year}{2017}\natexlab{}.
\newblock \showarticletitle{Mining User Reviews for Mobile App Comparisons}.
\newblock \bibinfo{journal}{\emph{Proceedings of the ACM on Interactive, Mobile, Wearable and Ubiquitous Technologies}} \bibinfo{volume}{1}, \bibinfo{number}{3} (\bibinfo{date}{Sept.} \bibinfo{year}{2017}), \bibinfo{pages}{75:1--75:15}.
\newblock


\bibitem[Lima et~al\mbox{.}(2023)]%
        {Lima2023}
\bibfield{author}{\bibinfo{person}{Vitor Lima}, \bibinfo{person}{Jacson Barbosa}, {and} \bibinfo{person}{Ricardo Marcacini}.} \bibinfo{year}{2023}\natexlab{}.
\newblock \bibinfo{title}{Learning Risk Factors from App Reviews: A Large Language Model Approach for Risk Matrix Construction}.
\newblock
\newblock
\urldef\tempurl%
\url{https://doi.org/10.21203/rs.3.rs-3182322/v1}
\showDOI{\tempurl}


\bibitem[Liu et~al\mbox{.}(2021)]%
        {LIU2021129}
\bibfield{author}{\bibinfo{person}{Huaxiao Liu}, \bibinfo{person}{Yihui Wang}, \bibinfo{person}{Yuzhou Liu}, {and} \bibinfo{person}{Shanquan Gao}.} \bibinfo{year}{2021}\natexlab{}.
\newblock \showarticletitle{Supporting features updating of apps by analyzing similar products in App stores}.
\newblock \bibinfo{journal}{\emph{Information Sciences}}  \bibinfo{volume}{580} (\bibinfo{year}{2021}), \bibinfo{pages}{129--151}.
\newblock
\showISSN{0020-0255}
\urldef\tempurl%
\url{https://doi.org/10.1016/j.ins.2021.08.050}
\showDOI{\tempurl}


\bibitem[Liu et~al\mbox{.}(2024)]%
        {Liu_2024}
\bibfield{author}{\bibinfo{person}{Z. Liu}, \bibinfo{person}{C. Chen}, \bibinfo{person}{J. Wang}, \bibinfo{person}{M. Chen}, \bibinfo{person}{B. Wu}, \bibinfo{person}{Z. Tian}, \bibinfo{person}{Y. Huang}, \bibinfo{person}{J. Hu}, {and} \bibinfo{person}{Q. Wang}.} \bibinfo{year}{2024}\natexlab{}.
\newblock \showarticletitle{Testing the Limits: Unusual Text Inputs Generation for Mobile App Crash Detection with Large Language Model}. In \bibinfo{booktitle}{\emph{2024 IEEE/ACM 46th International Conference on Software Engineering (ICSE)}}. \bibinfo{publisher}{IEEE Computer Society}, \bibinfo{address}{Los Alamitos, CA, USA}, \bibinfo{pages}{1685--1696}.
\newblock
\showISSN{1558-1225}
\urldef\tempurl%
\url{https://doi.ieeecomputersociety.org/}
\showURL{%
\tempurl}


\bibitem[Maalej and Nabil(2015)]%
        {7320414}
\bibfield{author}{\bibinfo{person}{Walid Maalej} {and} \bibinfo{person}{Hadeer Nabil}.} \bibinfo{year}{2015}\natexlab{}.
\newblock \showarticletitle{Bug report, feature request, or simply praise? On automatically classifying app reviews}. In \bibinfo{booktitle}{\emph{2015 IEEE 23rd International Requirements Engineering Conference (RE)}}. \bibinfo{pages}{116--125}.
\newblock
\urldef\tempurl%
\url{https://doi.org/10.1109/RE.2015.7320414}
\showDOI{\tempurl}


\bibitem[Mialon et~al\mbox{.}(2023)]%
        {mialon2023augmented}
\bibfield{author}{\bibinfo{person}{Grégoire Mialon}, \bibinfo{person}{Roberto Dessì}, \bibinfo{person}{Maria Lomeli}, \bibinfo{person}{Christoforos Nalmpantis}, \bibinfo{person}{Ram Pasunuru}, \bibinfo{person}{Roberta Raileanu}, \bibinfo{person}{Baptiste Rozière}, \bibinfo{person}{Timo Schick}, \bibinfo{person}{Jane Dwivedi-Yu}, \bibinfo{person}{Asli Celikyilmaz}, \bibinfo{person}{Edouard Grave}, \bibinfo{person}{Yann LeCun}, {and} \bibinfo{person}{Thomas Scialom}.} \bibinfo{year}{2023}\natexlab{}.
\newblock \bibinfo{title}{Augmented Language Models: a Survey}.
\newblock
\newblock
\showeprint[arxiv]{2302.07842}~[cs.CL]


\bibitem[Minaee et~al\mbox{.}(2024)]%
        {minaee2024large}
\bibfield{author}{\bibinfo{person}{Shervin Minaee}, \bibinfo{person}{Tomas Mikolov}, \bibinfo{person}{Narjes Nikzad}, \bibinfo{person}{Meysam Chenaghlu}, \bibinfo{person}{Richard Socher}, \bibinfo{person}{Xavier Amatriain}, {and} \bibinfo{person}{Jianfeng Gao}.} \bibinfo{year}{2024}\natexlab{}.
\newblock \showarticletitle{Large language models: A survey}.
\newblock \bibinfo{journal}{\emph{arXiv preprint arXiv:2402.06196}} (\bibinfo{year}{2024}).
\newblock


\bibitem[Mishra et~al\mbox{.}(2019)]%
        {Mishra2019}
\bibfield{author}{\bibinfo{person}{P. Mishra}, \bibinfo{person}{U. Singh}, \bibinfo{person}{C.~M. Pandey}, \bibinfo{person}{P. Mishra}, {and} \bibinfo{person}{G. Pandey}.} \bibinfo{year}{2019}\natexlab{}.
\newblock \showarticletitle{Application of student's t-test, analysis of variance, and covariance}.
\newblock \bibinfo{journal}{\emph{Annals of Cardiac Anaesthesia}} \bibinfo{volume}{22}, \bibinfo{number}{4} (\bibinfo{date}{October--December} \bibinfo{year}{2019}), \bibinfo{pages}{407--411}.
\newblock
\urldef\tempurl%
\url{https://doi.org/10.4103/aca.ACA_94_19}
\showDOI{\tempurl}


\bibitem[Motger et~al\mbox{.}(2024)]%
        {Quim_2024}
\bibfield{author}{\bibinfo{person}{Quim Motger}, \bibinfo{person}{Xavier Franch}, \bibinfo{person}{Vincenzo Gervasi}, {and} \bibinfo{person}{Jordi Marco}.} \bibinfo{year}{2024}\natexlab{}.
\newblock \showarticletitle{Unveiling Competition Dynamics in Mobile App Markets Through User Reviews}. In \bibinfo{booktitle}{\emph{Requirements Engineering: Foundation for Software Quality}}, \bibfield{editor}{\bibinfo{person}{Daniel Mendez} {and} \bibinfo{person}{Ana Moreira}} (Eds.). \bibinfo{publisher}{Springer Nature Switzerland}, \bibinfo{address}{Cham}, \bibinfo{pages}{251--266}.
\newblock
\showISBNx{978-3-031-57327-9}


\bibitem[O’Flaherty(2020)]%
        {Flaherty_2020}
\bibfield{author}{\bibinfo{person}{Kate O’Flaherty}.} \bibinfo{year}{2020}\natexlab{}.
\newblock \bibinfo{title}{Zoom beats Microsoft Teams, google meet with game-changing new features}.
\newblock
\newblock
\urldef\tempurl%
\url{https://www.forbes.com/sites/kateoflahertyuk/2020/10/14/zoom-beats-microsoft-teams-google-meet-with-game-changing-new-features/}
\showURL{%
\tempurl}


\bibitem[{Pagano} and {Maalej}(2013)]%
        {6636712}
\bibfield{author}{\bibinfo{person}{D. {Pagano}} {and} \bibinfo{person}{W. {Maalej}}.} \bibinfo{year}{2013}\natexlab{}.
\newblock \showarticletitle{User feedback in the appstore: An empirical study}. In \bibinfo{booktitle}{\emph{2013 21st IEEE International Requirements Engineering Conference (RE)}}. \bibinfo{pages}{125--134}.
\newblock
\urldef\tempurl%
\url{https://doi.org/10.1109/RE.2013.6636712}
\showDOI{\tempurl}


\bibitem[Ramos(2003)]%
        {Ramos}
\bibfield{author}{\bibinfo{person}{Juan Ramos}.} \bibinfo{year}{2003}\natexlab{}.
\newblock \showarticletitle{Using {TF-IDF} to determine word relevance in document queries}. In \bibinfo{booktitle}{\emph{Proceedings of the 1st instructional Conference on Machine Learning}} \emph{(\bibinfo{series}{iCML '03})}. \bibinfo{pages}{1--4}.
\newblock


\bibitem[Rigg and Myle(2021)]%
        {Rigg_Myle_2021}
\bibfield{author}{\bibinfo{person}{Christian Rigg} {and} \bibinfo{person}{Nikshep Myle}.} \bibinfo{year}{2021}\natexlab{}.
\newblock \bibinfo{title}{Zoom Video Conferencing Service Review}.
\newblock
\newblock
\urldef\tempurl%
\url{https://www.techradar.com/reviews/zoom#section-zoom-features}
\showURL{%
\tempurl}


\bibitem[Roumeliotis et~al\mbox{.}(2024)]%
        {ROUMELIOTIS2024100056}
\bibfield{author}{\bibinfo{person}{Konstantinos~I. Roumeliotis}, \bibinfo{person}{Nikolaos~D. Tselikas}, {and} \bibinfo{person}{Dimitrios~K. Nasiopoulos}.} \bibinfo{year}{2024}\natexlab{}.
\newblock \showarticletitle{LLMs in e-commerce: A comparative analysis of GPT and LLaMA models in product review evaluation}.
\newblock \bibinfo{journal}{\emph{Natural Language Processing Journal}}  \bibinfo{volume}{6} (\bibinfo{year}{2024}), \bibinfo{pages}{100056}.
\newblock
\showISSN{2949-7191}
\urldef\tempurl%
\url{https://doi.org/10.1016/j.nlp.2024.100056}
\showDOI{\tempurl}


\bibitem[S{\"a}nger et~al\mbox{.}(2016)]%
        {sanger-etal-2016-scare}
\bibfield{author}{\bibinfo{person}{Mario S{\"a}nger}, \bibinfo{person}{Ulf Leser}, \bibinfo{person}{Steffen Kemmerer}, \bibinfo{person}{Peter Adolphs}, {and} \bibinfo{person}{Roman Klinger}.} \bibinfo{year}{2016}\natexlab{}.
\newblock \showarticletitle{{SCARE} ― The Sentiment Corpus of App Reviews with Fine-grained Annotations in {G}erman}. In \bibinfo{booktitle}{\emph{Proceedings of the Tenth International Conference on Language Resources and Evaluation ({LREC}'16)}}, \bibfield{editor}{\bibinfo{person}{Nicoletta Calzolari}, \bibinfo{person}{Khalid Choukri}, \bibinfo{person}{Thierry Declerck}, \bibinfo{person}{Sara Goggi}, \bibinfo{person}{Marko Grobelnik}, \bibinfo{person}{Bente Maegaard}, \bibinfo{person}{Joseph Mariani}, \bibinfo{person}{Helene Mazo}, \bibinfo{person}{Asuncion Moreno}, \bibinfo{person}{Jan Odijk}, {and} \bibinfo{person}{Stelios Piperidis}} (Eds.). \bibinfo{publisher}{European Language Resources Association (ELRA)}, \bibinfo{address}{Portoro{\v{z}}, Slovenia}, \bibinfo{pages}{1114--1121}.
\newblock
\urldef\tempurl%
\url{https://aclanthology.org/L16-1178}
\showURL{%
\tempurl}


\bibitem[Scalabrino et~al\mbox{.}(2019)]%
        {8057860}
\bibfield{author}{\bibinfo{person}{Simone Scalabrino}, \bibinfo{person}{Gabriele Bavota}, \bibinfo{person}{Barbara Russo}, \bibinfo{person}{Massimiliano~Di Penta}, {and} \bibinfo{person}{Rocco Oliveto}.} \bibinfo{year}{2019}\natexlab{}.
\newblock \showarticletitle{Listening to the Crowd for the Release Planning of Mobile Apps}.
\newblock \bibinfo{journal}{\emph{IEEE Transactions on Software Engineering}} \bibinfo{volume}{45}, \bibinfo{number}{1} (\bibinfo{year}{2019}), \bibinfo{pages}{68--86}.
\newblock
\urldef\tempurl%
\url{https://doi.org/10.1109/TSE.2017.2759112}
\showDOI{\tempurl}


\bibitem[Shah et~al\mbox{.}(2016)]%
        {DBLP:conf/sigsoft/ShahSP16}
\bibfield{author}{\bibinfo{person}{Faiz~Ali Shah}, \bibinfo{person}{Yevhenii Sabanin}, {and} \bibinfo{person}{Dietmar Pfahl}.} \bibinfo{year}{2016}\natexlab{}.
\newblock \showarticletitle{Feature-based evaluation of competing apps}. In \bibinfo{booktitle}{\emph{Proceedings of the {ACM} International Workshop on App Market Analytics}} \emph{(\bibinfo{series}{WAMA '16})}. \bibinfo{pages}{15--21}.
\newblock


\bibitem[Shah et~al\mbox{.}(2018)]%
        {DBLP:journals/corr/abs-1810-05187}
\bibfield{author}{\bibinfo{person}{Faiz~Ali Shah}, \bibinfo{person}{Kairit Sirts}, {and} \bibinfo{person}{Dietmar Pfahl}.} \bibinfo{year}{2018}\natexlab{}.
\newblock \showarticletitle{The Impact of Annotation Guidelines and Annotated Data on Extracting App Features from App Reviews}.
\newblock \bibinfo{journal}{\emph{CoRR}}  \bibinfo{volume}{abs/1810.05187} (\bibinfo{year}{2018}).
\newblock


\bibitem[Shah et~al\mbox{.}(2019a)]%
        {DBLP:conf/refsq/ShahSP19}
\bibfield{author}{\bibinfo{person}{Faiz~Ali Shah}, \bibinfo{person}{Kairit Sirts}, {and} \bibinfo{person}{Dietmar Pfahl}.} \bibinfo{year}{2019}\natexlab{a}.
\newblock \showarticletitle{Is the {SAFE} Approach Too Simple for App Feature Extraction? {A} Replication Study}. In \bibinfo{booktitle}{\emph{Proceedings of the 25th International Working Conference on Requirements Engineering: Foundation for Software Quality}} \emph{(\bibinfo{series}{REFSQ 19})}. \bibinfo{pages}{21--36}.
\newblock


\bibitem[Shah et~al\mbox{.}(2019b)]%
        {ShahSP19}
\bibfield{author}{\bibinfo{person}{Faiz~Ali Shah}, \bibinfo{person}{Kairit Sirts}, {and} \bibinfo{person}{Dietmar Pfahl}.} \bibinfo{year}{2019}\natexlab{b}.
\newblock \showarticletitle{Using app reviews for competitive analysis: tool support}. In \bibinfo{booktitle}{\emph{Proceedings of the 3rd {ACM} {SIGSOFT} International Workshop on App Market Analytics}} \emph{(\bibinfo{series}{WAMA '19})}. \bibinfo{pages}{40--46}.
\newblock


\bibitem[Stokel-Walker(2020)]%
        {Stokel_Walker_2020}
\bibfield{author}{\bibinfo{person}{Chris Stokel-Walker}.} \bibinfo{year}{2020}\natexlab{}.
\newblock \bibinfo{title}{How Skype lost its crown to Zoom}.
\newblock
\newblock
\urldef\tempurl%
\url{https://www.wired.com/story/skype-coronavirus-pandemic/}
\showURL{%
\tempurl}


\bibitem[Strobelt et~al\mbox{.}(2023)]%
        {9908590}
\bibfield{author}{\bibinfo{person}{H. Strobelt}, \bibinfo{person}{A. Webson}, \bibinfo{person}{V. Sanh}, \bibinfo{person}{B. Hoover}, \bibinfo{person}{J. Beyer}, \bibinfo{person}{H. Pfister}, {and} \bibinfo{person}{A.~M. Rush}.} \bibinfo{year}{2023}\natexlab{}.
\newblock \showarticletitle{Interactive and Visual Prompt Engineering for Ad-hoc Task Adaptation with Large Language Models}.
\newblock \bibinfo{journal}{\emph{IEEE Transactions on Visualization and Computer Graphics}} \bibinfo{volume}{29}, \bibinfo{number}{01} (\bibinfo{date}{jan} \bibinfo{year}{2023}), \bibinfo{pages}{1146--1156}.
\newblock
\showISSN{1941-0506}
\urldef\tempurl%
\url{https://doi.org/10.1109/TVCG.2022.3209479}
\showDOI{\tempurl}


\bibitem[Su et~al\mbox{.}(2019)]%
        {9066126}
\bibfield{author}{\bibinfo{person}{Yanqi Su}, \bibinfo{person}{Yongchao Wang}, {and} \bibinfo{person}{Wenhua Yang}.} \bibinfo{year}{2019}\natexlab{}.
\newblock \showarticletitle{Mining and Comparing User Reviews across Similar Mobile Apps}. In \bibinfo{booktitle}{\emph{2019 15th International Conference on Mobile Ad-Hoc and Sensor Networks (MSN)}}. \bibinfo{pages}{338--342}.
\newblock
\urldef\tempurl%
\url{https://doi.org/10.1109/MSN48538.2019.00070}
\showDOI{\tempurl}


\bibitem[Tushev et~al\mbox{.}(2022)]%
        {9794110}
\bibfield{author}{\bibinfo{person}{M. Tushev}, \bibinfo{person}{F. Ebrahimi}, {and} \bibinfo{person}{A. Mahmoud}.} \bibinfo{year}{2022}\natexlab{}.
\newblock \showarticletitle{Domain-Specific Analysis of Mobile App Reviews Using Keyword-Assisted Topic Models}. In \bibinfo{booktitle}{\emph{2022 IEEE/ACM 44th International Conference on Software Engineering (ICSE)}}. \bibinfo{publisher}{IEEE Computer Society}, \bibinfo{address}{Los Alamitos, CA, USA}, \bibinfo{pages}{762--773}.
\newblock
\urldef\tempurl%
\url{https://doi.org/10.1145/3510003.3510201}
\showDOI{\tempurl}


\bibitem[Vu et~al\mbox{.}(2015)]%
        {DBLP:conf/kbse/VuNPN15}
\bibfield{author}{\bibinfo{person}{Phong~Minh Vu}, \bibinfo{person}{Tam~The Nguyen}, \bibinfo{person}{Hung~Viet Pham}, {and} \bibinfo{person}{Tung~Thanh Nguyen}.} \bibinfo{year}{2015}\natexlab{}.
\newblock \showarticletitle{Mining User Opinions in Mobile App Reviews: {A} Keyword-Based Approach {(T)}}. In \bibinfo{booktitle}{\emph{Proceedings of the 30th {IEEE/ACM} International Conference on Automated Software Engineering}} \emph{(\bibinfo{series}{ASE '15})}. \bibinfo{pages}{749--759}.
\newblock


\bibitem[Wang et~al\mbox{.}(2022a)]%
        {WANG2022111435}
\bibfield{author}{\bibinfo{person}{Yihui Wang}, \bibinfo{person}{Shanquan Gao}, \bibinfo{person}{Xingtong Li}, \bibinfo{person}{Lei Liu}, {and} \bibinfo{person}{Huaxiao Liu}.} \bibinfo{year}{2022}\natexlab{a}.
\newblock \showarticletitle{Missing standard features compared with similar apps? A feature recommendation method based on the knowledge from user interface}.
\newblock \bibinfo{journal}{\emph{Journal of Systems and Software}}  \bibinfo{volume}{193} (\bibinfo{year}{2022}), \bibinfo{pages}{111435}.
\newblock
\showISSN{0164-1212}
\urldef\tempurl%
\url{https://doi.org/10.1016/j.jss.2022.111435}
\showDOI{\tempurl}


\bibitem[Wang et~al\mbox{.}(2022b)]%
        {WANG2022118095}
\bibfield{author}{\bibinfo{person}{Yihui Wang}, \bibinfo{person}{Shanquan Gao}, \bibinfo{person}{Yan Zhang}, \bibinfo{person}{Huaxiao Liu}, {and} \bibinfo{person}{Yiran Cao}.} \bibinfo{year}{2022}\natexlab{b}.
\newblock \showarticletitle{UISMiner: Mining UI suggestions from user reviews}.
\newblock \bibinfo{journal}{\emph{Expert Systems with Applications}}  \bibinfo{volume}{208} (\bibinfo{year}{2022}), \bibinfo{pages}{118095}.
\newblock
\showISSN{0957-4174}
\urldef\tempurl%
\url{https://doi.org/10.1016/j.eswa.2022.118095}
\showDOI{\tempurl}


\bibitem[Wei(2023)]%
        {wei2023enhancing}
\bibfield{author}{\bibinfo{person}{Jialiang Wei}.} \bibinfo{year}{2023}\natexlab{}.
\newblock \showarticletitle{Enhancing Requirements Elicitation through App Stores Mining: Health Monitoring App Case Study}. In \bibinfo{booktitle}{\emph{2023 IEEE 31st International Requirements Engineering Conference (RE)}}. IEEE, \bibinfo{pages}{396--400}.
\newblock


\bibitem[Wei et~al\mbox{.}(2023)]%
        {10356483}
\bibfield{author}{\bibinfo{person}{Jialiang Wei}, \bibinfo{person}{Anne-Lise Courbis}, \bibinfo{person}{Thomas Lambolais}, \bibinfo{person}{Binbin Xu}, \bibinfo{person}{Pierre~Louis Bernard}, {and} \bibinfo{person}{Gérard Dray}.} \bibinfo{year}{2023}\natexlab{}.
\newblock \showarticletitle{Zero-shot Bilingual App Reviews Mining with Large Language Models}. In \bibinfo{booktitle}{\emph{2023 IEEE 35th International Conference on Tools with Artificial Intelligence (ICTAI)}}. \bibinfo{pages}{898--904}.
\newblock
\urldef\tempurl%
\url{https://doi.org/10.1109/ICTAI59109.2023.00135}
\showDOI{\tempurl}


\bibitem[Xu et~al\mbox{.}(2023)]%
        {xu2023limits}
\bibfield{author}{\bibinfo{person}{Xiancai Xu}, \bibinfo{person}{Jia-Dong Zhang}, \bibinfo{person}{Rongchang Xiao}, {and} \bibinfo{person}{Lei Xiong}.} \bibinfo{year}{2023}\natexlab{}.
\newblock \bibinfo{title}{The Limits of ChatGPT in Extracting Aspect-Category-Opinion-Sentiment Quadruples: A Comparative Analysis}.
\newblock
\newblock
\showeprint[arxiv]{2310.06502}~[cs.CL]


\bibitem[Xue et~al\mbox{.}(2024)]%
        {xue2024automated}
\bibfield{author}{\bibinfo{person}{Pengyu Xue}, \bibinfo{person}{Linhao Wu}, \bibinfo{person}{Zhongxing Yu}, \bibinfo{person}{Zhi Jin}, \bibinfo{person}{Zhen Yang}, \bibinfo{person}{Xinyi Li}, \bibinfo{person}{Zhenyu Yang}, {and} \bibinfo{person}{Yue Tan}.} \bibinfo{year}{2024}\natexlab{}.
\newblock \showarticletitle{Automated Commit Message Generation with Large Language Models: An Empirical Study and Beyond}.
\newblock \bibinfo{journal}{\emph{arXiv preprint arXiv:2404.14824}} (\bibinfo{year}{2024}).
\newblock


\bibitem[Zhang et~al\mbox{.}(2023)]%
        {zhang2023sentiment}
\bibfield{author}{\bibinfo{person}{Wenxuan Zhang}, \bibinfo{person}{Yue Deng}, \bibinfo{person}{Bing Liu}, \bibinfo{person}{Sinno~Jialin Pan}, {and} \bibinfo{person}{Lidong Bing}.} \bibinfo{year}{2023}\natexlab{}.
\newblock \bibinfo{title}{Sentiment Analysis in the Era of Large Language Models: A Reality Check}.
\newblock
\newblock
\showeprint[arxiv]{2305.15005}~[cs.CL]


\bibitem[Zhao et~al\mbox{.}(2023)]%
        {zhao2023survey}
\bibfield{author}{\bibinfo{person}{Wayne~Xin Zhao}, \bibinfo{person}{Kun Zhou}, \bibinfo{person}{Junyi Li}, \bibinfo{person}{Tianyi Tang}, \bibinfo{person}{Xiaolei Wang}, \bibinfo{person}{Yupeng Hou}, \bibinfo{person}{Yingqian Min}, \bibinfo{person}{Beichen Zhang}, \bibinfo{person}{Junjie Zhang}, \bibinfo{person}{Zican Dong}, \bibinfo{person}{Yifan Du}, \bibinfo{person}{Chen Yang}, \bibinfo{person}{Yushuo Chen}, \bibinfo{person}{Zhipeng Chen}, \bibinfo{person}{Jinhao Jiang}, \bibinfo{person}{Ruiyang Ren}, \bibinfo{person}{Yifan Li}, \bibinfo{person}{Xinyu Tang}, \bibinfo{person}{Zikang Liu}, \bibinfo{person}{Peiyu Liu}, \bibinfo{person}{Jian-Yun Nie}, {and} \bibinfo{person}{Ji-Rong Wen}.} \bibinfo{year}{2023}\natexlab{}.
\newblock \bibinfo{title}{A Survey of Large Language Models}.
\newblock
\newblock
\showeprint[arxiv]{2303.18223}~[cs.CL]


\bibitem[Zhou et~al\mbox{.}(2023)]%
        {zhou2023large}
\bibfield{author}{\bibinfo{person}{Yongchao Zhou}, \bibinfo{person}{Andrei~Ioan Muresanu}, \bibinfo{person}{Ziwen Han}, \bibinfo{person}{Keiran Paster}, \bibinfo{person}{Silviu Pitis}, \bibinfo{person}{Harris Chan}, {and} \bibinfo{person}{Jimmy Ba}.} \bibinfo{year}{2023}\natexlab{}.
\newblock \showarticletitle{Large Language Models are Human-Level Prompt Engineers}. In \bibinfo{booktitle}{\emph{The Eleventh International Conference on Learning Representations}}.
\newblock
\urldef\tempurl%
\url{https://openreview.net/forum?id=92gvk82DE-}
\showURL{%
\tempurl}


\end{thebibliography}

\appendix

\end{document}